\documentclass[prl,aps,showpacs,twocolumn,preprintnumbers,amsmath,amssymb,superscriptaddress]{revtex4-1}
\usepackage{amsmath,amssymb,amsfonts}
\usepackage{graphicx}
\usepackage[colorlinks=True,linkcolor=red,citecolor=blue,urlcolor=blue]{hyperref}
\usepackage{xcolor}
\usepackage{chngcntr}
\usepackage{braket}
\usepackage{hyperref}
\usepackage{nicefrac}
\usepackage{bm}

\begin{document}
\bibliographystyle{apsrev4-1}
\title{Difference frequency generation in topological semimetals}

\author{F. de Juan}
\affiliation{Donostia International Physics Center, P. Manuel de Lardizabal 4, 20018 Donostia-San Sebastian, Spain}
\affiliation{IKERBASQUE, Basque Foundation for Science, Maria Diaz de Haro 3, 48013 Bilbao, Spain}
\author{Y. Zhang}
\affiliation{Department of Physics, Massachusetts Institute of Technology, Cambridge, Massachusetts 02139, USA}
\author{T. Morimoto}
\affiliation{Department of Applied Physics, The University of Tokyo, Tokyo, 113-8656, Japan}
\author{Y. Sun}
\affiliation{Max Planck Institute for Chemical Physics of Solids, 01187 Dresden, Germany}
\author{Joel E. Moore}
\affiliation{Department of Physics, University of California, Berkeley, CA 94720, USA}
\author{ A. G. Grushin}
\affiliation{Institut Ne\'el, CNRS and Universit\'e Grenoble Alpes, Grenoble, France}

\date{\today}
\begin{abstract}

When two lasers are applied to a non-centrosymmetric material, light can be generated at the difference of the incoming frequencies $\Delta\omega$, a phenomenon known as difference frequency generation (DFG), well characterized in semiconductors. In this work, we derive a general expression for DFG in metals, which we use to show that the DFG in chiral topological semimetals under circular polarized light is quantized in units of $e^3/h^2$ and independent of material parameters, including the scattering time $\tau$, when $\Delta\omega \gg \tau^{-1}$. In this regime, DFG provides a simpler alternative to measure a quantized response in metals compared to previous proposals based on single frequency experiments. Our general derivation unmasks, in addition, a free-carrier contribution to the circular DFG beyond the semiclassical one. This contribution can be written as a Fermi surface integral, features strong frequency dependence, and oscillates with a $\pi/2$ shift with respect to the quantized contribution. We make predictions for the circular DFG of chiral and non-chiral materials using generic effective models, and ab-initio calculations for TaAs and RhSi. Our work provides a complete picture of the DFG in the length gauge approach, in the clean, non-interacting limit, and highlights a plausible experiment to measure topologically quantizated photocurrents in metals.
\end{abstract}
\maketitle

One of the most striking predictions that follow from the protected point-like band crossings in topological semimetals is the first quantized observable defined for a metal. It is a circular photocurrent that grows in time at a universal rate given by fundamental constants only in chiral topological semimetals~\cite{deJuan17,Flicker2018}, which lack mirror symmetries~\cite{HOI15,HXB16,KramersWeyl}. This steadily growing current, known as injection current, originates from resonant, interband transitions and is proportional to the intensity of light. Its universal rate of growth, given by the trace of the circular photogalvanic tensor $\beta^{ab}$, cannot be directly extracted from a steady-state experiment because at times longer than the scattering time $t\gg \tau$, the current saturates to a $\tau$-dependent value, which makes the  measurement of the quantized response challenging. 

To measure the intrinsic quantized current rate it is rather desirable to use time-dependent electric fields in the form of light pulses of duration shorter than $\tau$, and measure the emitted THz fields. However, typical experiments are performed in the opposite regime~\cite{Laman05,Bieler06,BMB16}, where assumptions on the nature of scattering are required to extract the injection current~\cite{Rees19}. 

In this work, we propose an alternative to access the quantized current rate, which is to measure difference frequency generation (DFG), where two monochromatic light beams of frequencies $\omega \pm \Delta \omega/2$ produce a slowly oscillating current of frequency $\Delta \omega$. In the limit $\Delta \omega \ll \omega$, DFG is formally equivalent to a photogalvanic effect, but if in addition we demand that $\Delta \omega \gg \tau^{-1}$, the response is intrinsic and $\tau$ independent, exposing the universal quantum. 

To show this, we calculate the circular DFG response for all metals in the regime $\omega \gg \Delta \omega \gg \tau^{-1}$, using the length gauge formalism~\cite{GhahramaniSipe,AversaSipe,SipeShkrebtii,NastosSipe06,NastosSipe10}. We find that the interband contribution to circular DFG oscillates exactly out of phase with respect to the incoming light, and is given by the intrinsic injection rate $\beta^{ab}$, becoming topologically quantized and independent of material parameters in chiral topological metals. In addition, we find that there is a free-carrier contribution to circular DFG~\cite{Genkin68} for any metal which oscillates in phase with the incoming light and displays strong $\omega$ dependence due to an extra term beyond the semiclassical Berry-dipole~\cite{sodemannfu}. Due to this additional term, which reduces to a universal function for linear band crossings at time-reversal invariant momenta, the total circular DFG in metals can have any phase shift compared to the incoming light. We present predictions for both interband and free-carrier contributions to the circular DFG using several models for topological semimetals as well as ab-initio calculations. Finally, we briefly discuss the subtleties of the free-carrier DC photocurrents in the opposite limit $\Delta \omega \ll \tau^{-1}$~\cite{BelinicherTransient,BelinicherKinetic}. Our work contributes to elucidate how topological properties, responsible for unusual non-linear effects in topological semimetals~\cite{Ran17,Zhang18,Ma17,Burch17,Wu17,Patankar2018,Rappe18,Hosur11,MN16,Parker18}, determine non-monochromatic responses.

\emph{Difference frequency generation in metals} - DFG is the current response obtained when shining two monochromatic beams  
$\mathcal{E}_{i}^a(t) = {\rm Re} [E^a_i e^{-i\omega_i t}]$ with $i=1,2$ with average frequency $\omega = (\omega_1 + \omega_2)/2$ and difference $\Delta \omega = \omega_1 - \omega_2$. For concreteness we consider equal polarizations $E^a_1= E^a_2=E^a$, and assume time-reversal symmetry throughout the manuscript (we consider the general case in the Supplemental Material~\cite{Supplementalphotoweyl}) . In the mentioned regime $\omega \gg \Delta \omega \gg \tau^{-1}$, the generated DFG current can be expanded in perturbation theory in $\Delta \omega/\omega$ as
\begin{align}
J^a(t)  =& 4\left[\frac{\sin(\Delta\omega t)}{\Delta\omega} \; \beta^{ab}(\omega) +  \cos (\Delta \omega t) \;\gamma^{ab}(\omega)  \right] [\vec E \times \vec E^*]^b \nonumber \\
&+ 2\cos(\Delta \omega t) \; \sigma^{abc}(\omega)(E^bE^{c*} +E^cE^{b*}),
\label{eq:allDFG}\end{align}
Defining $\gamma^{ab} = \gamma_1^{ab} + \gamma_2^{ab}$, the explicit expressions for these tensors are 
\begin{align}
{\beta^{ab}}(\omega) &= {\frac{i\pi C}{4}} \int_k {\sum_{n>m}} f_{nm} \Delta^a_{mn}{\rm Im} [r^d_{nm}r^c_{mn}] \epsilon^{bcd} \delta(\omega_{mn}-\omega) \label{eq:injection}\\
{\sigma^{abc}}(\omega) &= \frac{\pi C}{2} \int_k {\sum_{n>m}} f_{nm} {\rm Im} [r_{mn}^b r_{nm;a}^c]  \delta(\omega_{mn}-\omega)\label{lambda} \\
\gamma_1^{ab}(\omega) &= \frac{iC}{2} \int_k {\sum_{n>m}} \frac{\omega f_{nm,a}}{\omega_{nm}^2-\omega^2}  {\rm Im}[\epsilon^{bcd}r^d_{nm}r^c_{mn}]\label{FSintraband}\\
\gamma_2^{ab} (\omega)&= -\frac{iC}{2\omega} \int_k  \sum_n  \left[(f_{n,a} \Omega_n^b - \delta^{ab} f_{n,c}\Omega_n^c)\right]\label{eq:semicl}
\end{align}
where we assume $\omega>0$. In these equations $C=e^3/\hbar^2$, $\int_k =\int d^3k/(2\pi)^3$,  $f_n$ is the Fermi function which we take at zero temperature, $f_{nm} = f_n - f_m$, the Bloch eigenstates are $H \left|n\right> = \omega_n  \left|n\right>$, $\omega_{nm} = \omega_n -\omega_m$, $\xi^a_{nn} = i \left<n|\partial_{k_a}n\right>$ is the diagonal Berry connection, and $r_{nm}^a = i \left<n|\partial_{k_a}m\right>$ is the interband position matrix element, which is zero unless $n\neq m$. We represent derivatives with a comma, as in $f_{n,a} = \partial_{k_a} f_n$, while a semicolon denotes the generalized derivative, as in $r^b_{nm;a} = \partial_{k_a}r_{nm}^b -i(\xi^a_{nn}-\xi^a_{mm})r_{nm}^b$. The Berry curvature is $\Omega^a = \epsilon^{abc} \xi^c_{nn,b}$.

The tensors $\beta^{ab}$ and $\sigma^{abc}$ in Eq.~\eqref{eq:allDFG} are of interband origin and are the DFG analogs of the monochromatic photogalvanic effects known as injection and shift currents, respectively.  The tensor $\gamma^{ab}$ is a free-carrier, intraband contribution only present in metals. The focus of this work lies on $\beta^{ab}$ and $\gamma^{ab}$ which we label as circular DFG contributions since, unlike $\sigma^{abc}$, they lead to currents which change sign when the helicity of circularly polarized light is reversed.  

The interband contribution to the circular DFG, $\beta^{ab}$, has the same functional form as the interband circular photogalvanic effect. As anticipated, this implies that the trace of $\beta^{ab}$, and thus the corresponding DFG contribution Eq.~\eqref{eq:allDFG}, will be quantized in terms of fundamental constants in chiral topological semimetals~\cite{deJuan17,Flicker2018}.

The free-carrier contribution to the circular DFG, $\gamma^{ab}$, is closely related to the optical Hall conductivity $\sigma_{\rm Hall}^{ab}(\omega)$~ \cite{Steiner17}. As we show in~\cite{Supplementalphotoweyl}, $\gamma^{ad}\epsilon^{dbc}$ can be obtained from  $\sigma^{bc}_{\rm Hall}(\omega)$ by replacing $f_n \to f_{n,a}/\omega$ in its integrand, which explicitly reveals its metallic origin ~\cite{Genkin68}. 

Within $\gamma^{ab}$ the free-carrier part $\gamma_2^{ab}$ given by Eq.~\eqref{eq:semicl}, is the well studied Berry-dipole term of semiclassical origin and a Fermi surface property~\cite{Deyo09,MooreOrenstein,sodemannfu,Morimoto16,IHU16,Rostami17,ZSY18}. The novel contribution we report is $\gamma_1^{ab}$ in Eq.~\eqref{FSintraband}, which is also written as Fermi surface integral. Note that its integrand depends on all bands so $\gamma_1^{ab}$ is not a Fermi surface property. Unlike the $1/\omega$ dependence of the semiclassical contribution, $\gamma_1^{ab}$ displays a strong $\omega$ dependence away from zero frequency. At every frequency where a new interband transition becomes active, the energy denominator in Eq.~\eqref{FSintraband} vanishes and the response diverges. Interestingly, $\gamma_1^{ab}$ can also be written in terms of Kramers-Kronig transforms of resonant contributions, a useful property to calculate $\gamma_1^{ab}$ using ab-initio calculations (see Supplemental Material~\cite{Supplementalphotoweyl}).

\emph{DFG of a single Weyl node} - As the simplest example, we first present the circular DFG tensors for a general Weyl semimetal node of the form 
\begin{equation}
H = \sigma_i v_{ij} k_j + u_i k_i \label{Hnode},
\end{equation}
written in terms of the Fermi velocity matrix $v_{ij}$, and its tilt $u_i$ in the crystallographic coordinate system. We present the detailed computation of the tensors $\beta^{ab}$, $\gamma_1^{ab}$, $\gamma_2^{ab}$ in~\cite{Supplementalphotoweyl} and expressed their final form in terms of trace and trace-free parts as 
\begin{align}
\beta^{ab} = \chi \left(\frac{\delta^{ab}}{3}\beta+\left[\frac{\delta^{ab}}{3}-v_{ac} v^{-1}_{bd}\frac{\tilde{u}_c \tilde{u}_d}{\tilde{u}^2}\right]  \beta_F \right), \label{trdecomp}
\end{align}
where $\chi = \tfrac{{\rm det} v}{|{\rm det} v|} = \pm1$ is the chirality of the node, $v^{-1}_{ij} \equiv (v^{-1})_{ij}$ and $\tilde{u}_i = v^{-1}_{ij} u_j$, $\tilde{u} = \sqrt{\tilde{u}_i^2}$, and similar expressions hold for $\gamma_1^{ab}$ and $\gamma_2^{ab}$. In terms of the universal photogalvanic constant $\beta_0\equiv i C/4\pi=i\pi e^3/h^2$ the trace parts are 
\begin{align}
\beta &= -\tfrac{1}{8}\beta_0(1-g_1(\omega)), \label{trbeta} \\
\gamma_1 &= -\dfrac{\beta_0}{2\pi} \frac{-\tilde{u} \omega +\mu\; g_2(\omega)}{{\tilde{u}} \omega^2} \label{trgamma}, \hspace{0.5cm} 
\gamma_2 =  -\dfrac{\beta_0}{\pi\omega},
\end{align}
and the trace-free parts are
\begin{align}
\beta_F &= \tfrac{1}{8}\beta_0 g_1(\omega)(1-[g_1(\omega)]^2), \\
\gamma_{1,F} &= -\dfrac{\beta_0}{8\pi}\frac{1}{\tilde{u}^3 \omega^2}\Big[ (4\tilde{u}^3-6 \tilde{u})\omega^2 -  (4\tilde{u}^2-12)\mu \omega g_2(\omega) \Big. \nonumber \\
&\Big. + 3 (4 \mu^2 + (1 - \tilde{u}^2) \omega^2) g_3(\omega) \Big],\\
\gamma_{2,F} &= -\frac{\beta_0}{4\pi\omega \tilde{u}^3}\left[2\tilde{u}^3- 3(\tilde{u}+(\tilde{u}^2-1) {\rm arctanh}(\tilde{u})\right] ,
\end{align}
where $g_1(\omega) = a\Theta(1+a)\Theta(1-a) + \Theta(a-1)-\Theta(-a-1)$ with $a = (\tfrac{2\mu}{\omega}-1)/\tilde{u}$, $g_2(\omega) = {\rm arctanh}\left( \frac{4\tilde{u} \mu \omega}{4\mu^2+(\tilde{u}^2-1)\omega^2}\right)$ and $g_3(\omega) = {\rm arctanh}\left(\frac{2\tilde{u} \omega^2}{-4\mu^2+(\tilde{u}^2+1)\omega^2}\right)$.

\begin{figure}[t]
\includegraphics[width=\linewidth]{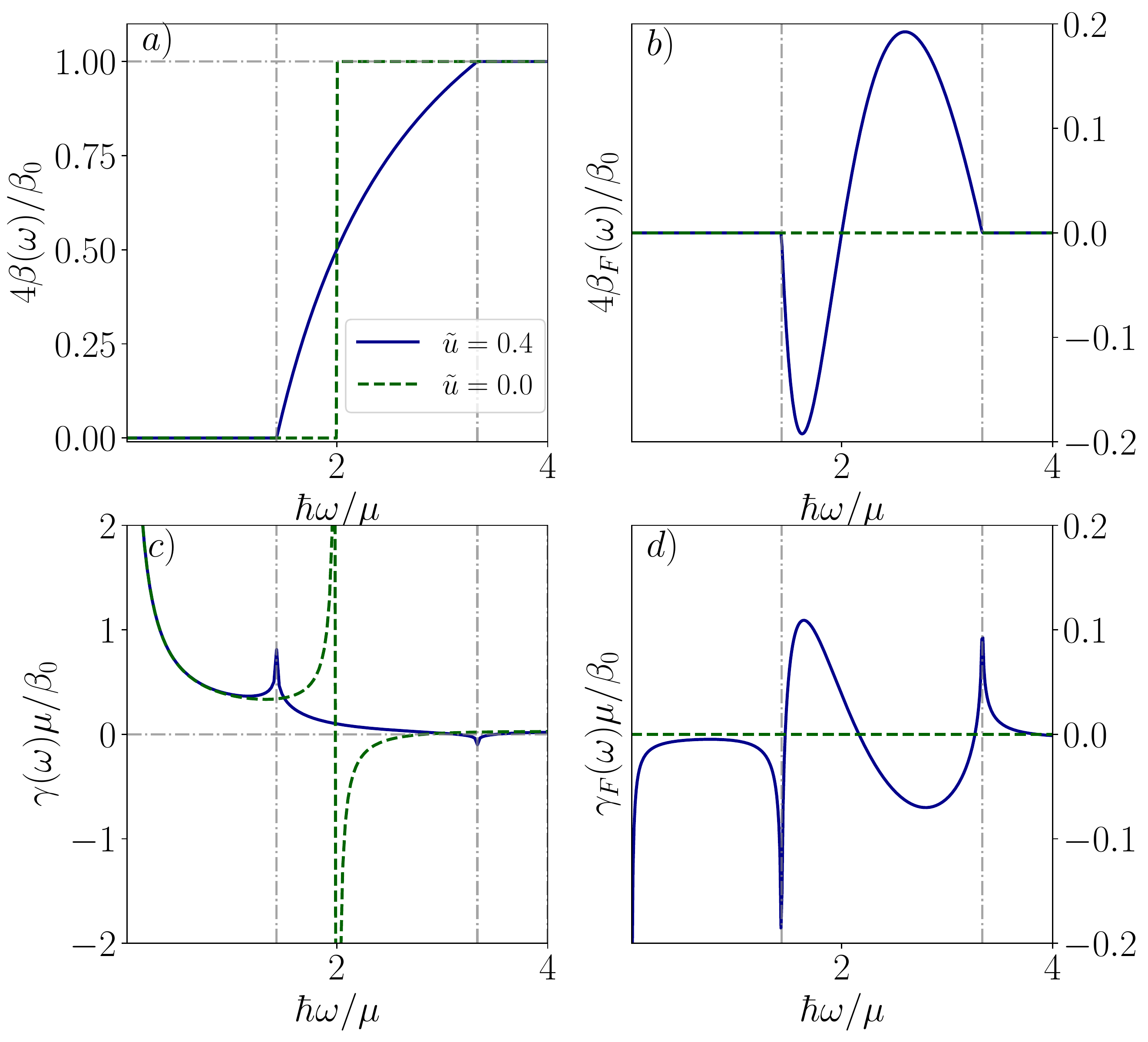}
\caption{Trace and traceless parts of CPGE tensors for a Weyl node with chirality $\chi=-1$, with tilt $\tilde{u} = 0$ (green dashed) and $\tilde{u} = 0.4$ (blue solid). a) Trace of injection tensor $\beta$, where a quantized plateau is observed. b) Traceless part of injection $\beta_F$.  c) Trace of free-carrier tensor $\gamma$. Note the $1/\omega$ divergence stemming from $\gamma_2$ which always cancels after summing over all nodes in the Brillouin zone. d) Traceless part of free-carrier $\gamma_F$. a) b) and c) d) are in units of $\beta_0=i\pi e^3/h^2$ and $\beta_0/\mu$, respectively. Vertical dashed lines mark the frequencies $\hbar\omega_\pm = 2\mu/(1\pm \tilde{u})$.}
\label{fig1}
\end{figure}

The trace ($\beta$) and traceless ($\beta_F$) parts are plotted in Fig.~\ref{fig1} a) and b) for zero and finite tilt. 
As shown in Ref.~\cite{deJuan17},  $\beta$ displays a quantized plateau once the resonant manifold of optical transitions becomes closed, because it is then determined by the monopole charge of the node. In the tilted case, $\tilde{u}\neq 0$ this manifold is open for $2\mu/(1+\tilde{u})< \omega < 2\mu/(1-\tilde{u})$, and $\beta$ becomes quantized for $ \omega > 2\mu/(1-\tilde{u})$. $\beta_F$ is finite only for $2\mu/(1+\tilde{u})< \omega < 2\mu/(1-\tilde{u})$, and vanishes in the zero tilt limit. Off-diagonal components of $\beta^{ab}$ therefore require finite tilt~\cite{Chan16,Konig17,Golub17}. 

The free-carrier trace ($\gamma = \gamma_1 + \gamma_2$) and traceless ($\gamma_F = \gamma_{1,F}+\gamma_{2,F}$) parts are shown in Fig.~\ref{fig1} c) and d). In the zero tilt limit, $\gamma_1 \propto \omega/(4\mu^2-\omega^2)$. At finite tilt, the divergence at $2\mu$ splits into two logarithmic divergences at $\omega_{\pm} = 2\mu/(1\pm \tilde{u})$, where $\gamma_{1,F}$ also displays strong singularities. We note that $\gamma_1^{ab}$ is well behaved when $\omega\to 0$.
The singularities at $\omega\to 0$ originate from $\gamma_2$ and $\gamma_{2,F}$. While 
the trace of the semiclassical free-carrier contribution $\gamma_2$ gives a universal $1/\omega$ divergence that is independent of the chemical potential and tilt~\cite{Rostami17}, $\gamma_{2,F}$ displays a non-universal $1/\omega$ divergence, which depends on the tilt. Since in any material the number of left and right chiralities must be equal, the sum over Weyl nodes will cancel all the trace contributions $\gamma_2$ in pairs. The total free-carrier current, however, may have a non-universal $1/\omega$ pole due to $\gamma_{2,F}$ because of its tilt dependence and the tensor structure in Eq.~\eqref{trdecomp}.

\emph{DFG quantization in chiral topological semimetals} - 
For the trace $\beta$ to be quantized, a topological semimetal is required to have left and right chirality nodes at different energies~\cite{deJuan17,Flicker2018}, which is only allowed in a chiral lattice structure, which lacks mirror symmetries.

\begin{figure}[t!]
\includegraphics[width=\linewidth]{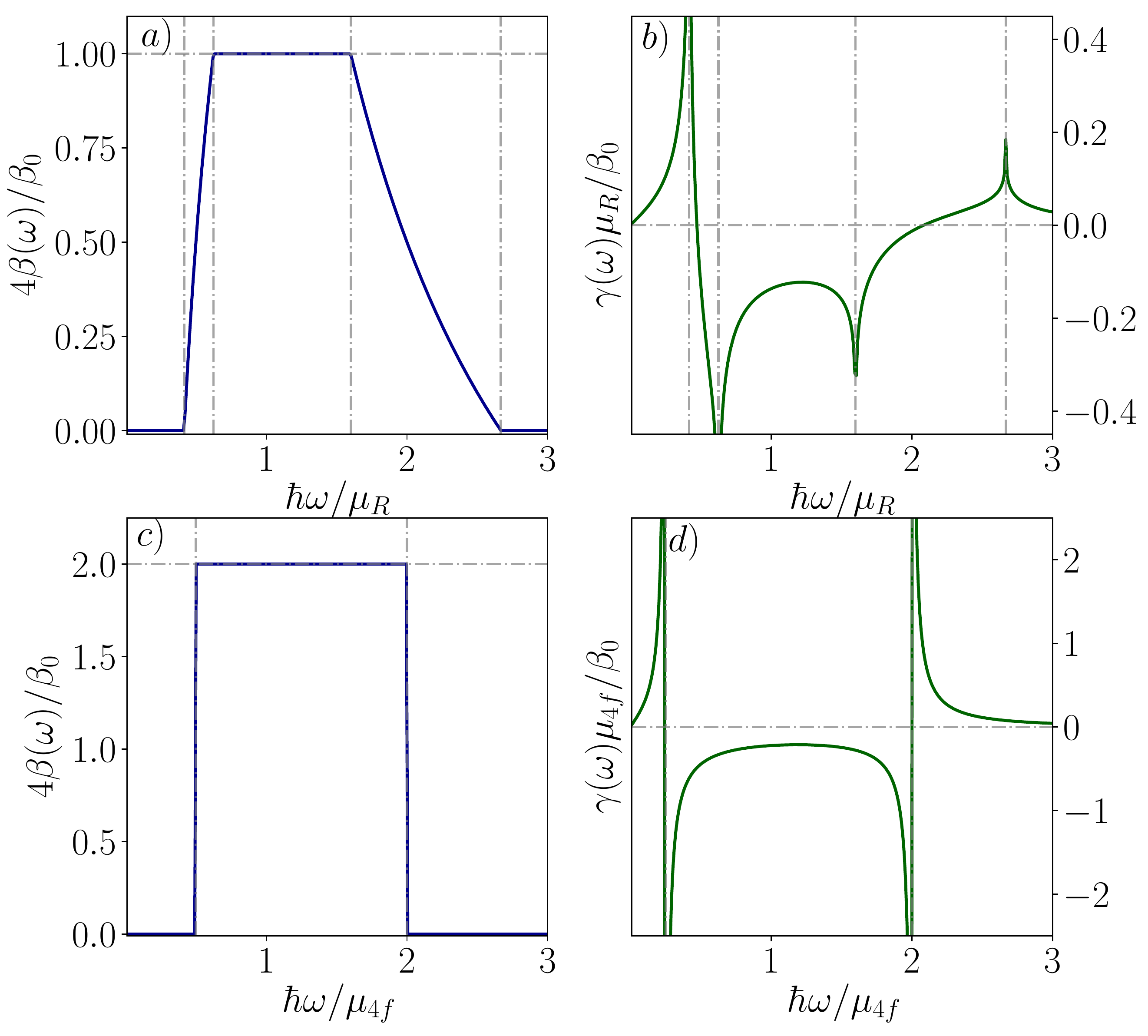}
\caption{a) and b): Diagonal parts of the circular DFG, $\beta$ (blue) and $\gamma$ (green) for a chiral Weyl semimetal with two Weyl nodes at energies $\mu_L$ and $\mu_R$, with $\mu_L/\mu_R=0.25$, and tilts $\tilde{u}_L=0.2$ $\tilde{u}_R=0.25$. c) and d): Same quantities for the linear approximation of a multifold material in space group 198, with a threefold at $\mu_{3f}$ and a double Weyl fourfold fermion at $\mu_{4f}$ with $\mu_{3f}/\mu_{4f}=0.25$.}
\label{fig2}
\end{figure}

Consider first two Weyl nodes of opposite chiralities away from time reversal invariant points located at energies $\mu_L$ and $\mu_R$ measured from the chemical potential and tilts $\tilde{u}_L$ and $\tilde{u}_R$ (e.g. SrSi$_2$ without spin-orbit coupling~\cite{HXB16}). In the presence of time-reversal symmetry two more symmetry related nodes exist, which contribute in exactly the same way and simply double the result we present. 
In Fig.~\ref{fig2} a) and b) we show the circular DFG trace parts $\beta$ and $\gamma$.
The quantized plateau seen for $\beta$ is realized in the range $ 2\mu_L/(1 - \tilde{u}_L) < \omega < 2\mu_R/(1 + \tilde{u}_R)$, which determines how large the tilts can be before quantization is lost. 
Note also that, upon summing over the two chiralities, the total trace-part $\gamma$ presents no $1/\omega$ pole, as discussed above. 

Our second example are multifold fermions. They are low energy excitations close to degeneracy nodes where three, four, or six bands meet, and which requires additional crystalline symmetries to remain degenerate~\cite{manes2012,BradlynEA17,ChangEA17,TangEA17,Bouhon2017}. They have a definite chirality and larger monopole charge compared to Weyl nodes. In space group 198, to linear order in momentum and neglecting spin-orbit coupling, a threefold fermion exists at the $\Gamma$ point with a Hamiltonian $H = v_F\mathbf{k}\cdot\mathbf{S}$, where $\mathbf{S}$ is a vector of three spin-1 matrices. At the $R$ point, a fourfold fermion exists composed of two Weyl nodes of equal chiralities, $H = -v_F\mathbf{k}\cdot\mathbf{\sigma}\otimes 1$, separated in energy from the threefold at $\Gamma$. This situation is realized in CoSi~\cite{Takane2019,Rao:2019uw,Sanchez:2019wl}, RhSi~\cite{Sanchez:2019wl}, and AlPt~\cite{Schroter:2019kf}, which motivate our example.

In Fig.~\ref{fig2} c) and d) we show the circular DFG trace parts $\beta$ and $\gamma$ corresponding to a threefold and a fourfold nodes at energies $\mu_{3f}$ and $\mu_{4f}$ (see the Supplemental Material~\cite{Supplementalphotoweyl} for analytic expressions). Due to the monopole charge carried by the multifold fermions~\cite{Flicker2018}, $\beta$ displays a quantized plateau at $\beta_0/2$, two times that of a Weyl node. The free-carrier part $\gamma$ displays divergences at every energy where a new band becomes resonant with the Fermi level. As for the chiral Weyl case, upon summing over chiralities the low-energy divergent semiclassical part is absent. The remaining free-carrier contribution is $\gamma_1= (\beta_0/\pi) \sum_{i}\chi_i\omega/(\mu_i^2-\omega^2)$ where $\mu_i=\mu_{3f},2\mu_{4f}$ and $\chi_i=\pm$. This is a universal function for linear nodal points occurring at time-reversal invariant momenta.

Similarly to multifods, Kramers Weyl~\cite{KramersWeyl} (e.g. occurring in elemental Te or TlTe$_2$O$_6$) present Weyl nodes at time-reversal invariant momenta separated in energy. The full DFG tensor can be calculated analytically, even with quadratic corrections, and it is detailed in the Supplemental Material~\cite{Supplementalphotoweyl} since it is conceptually similar to our previous examples.

\emph{DFG in TaAs and RhSi from first principles} -  
In the presence of mirror symmetries the trace of $\gamma^{ab}$ and $\beta^{ab}$ vanish, with the consequent loss of quantization.
In a system with $C_{4v}$ symmetry like TaAs~\cite{Weng2015,Huang:vn,LvXu2015,Xu613,Yang:2015ev}, the only allowed component is antisymmetric $\beta^{xy}=-\beta^{yx}$, which can only be due to the traceless part in Eq.~\eqref{trdecomp}.
The same requirement holds for $\gamma^{ab}$.
Moreover, from Eq.~\eqref{trdecomp} both tilt and non-trivial $v_{ij}$ are required to produce a finite result in TaAs. 
If $v_{ij} = \delta_{ij}$ this component must also be zero~\cite{Golub17,Golub19}, regardless of the tilt, since Eq.~\eqref{trdecomp} becomes symmetric. 

\begin{figure}[t!]
\includegraphics[width=\linewidth]{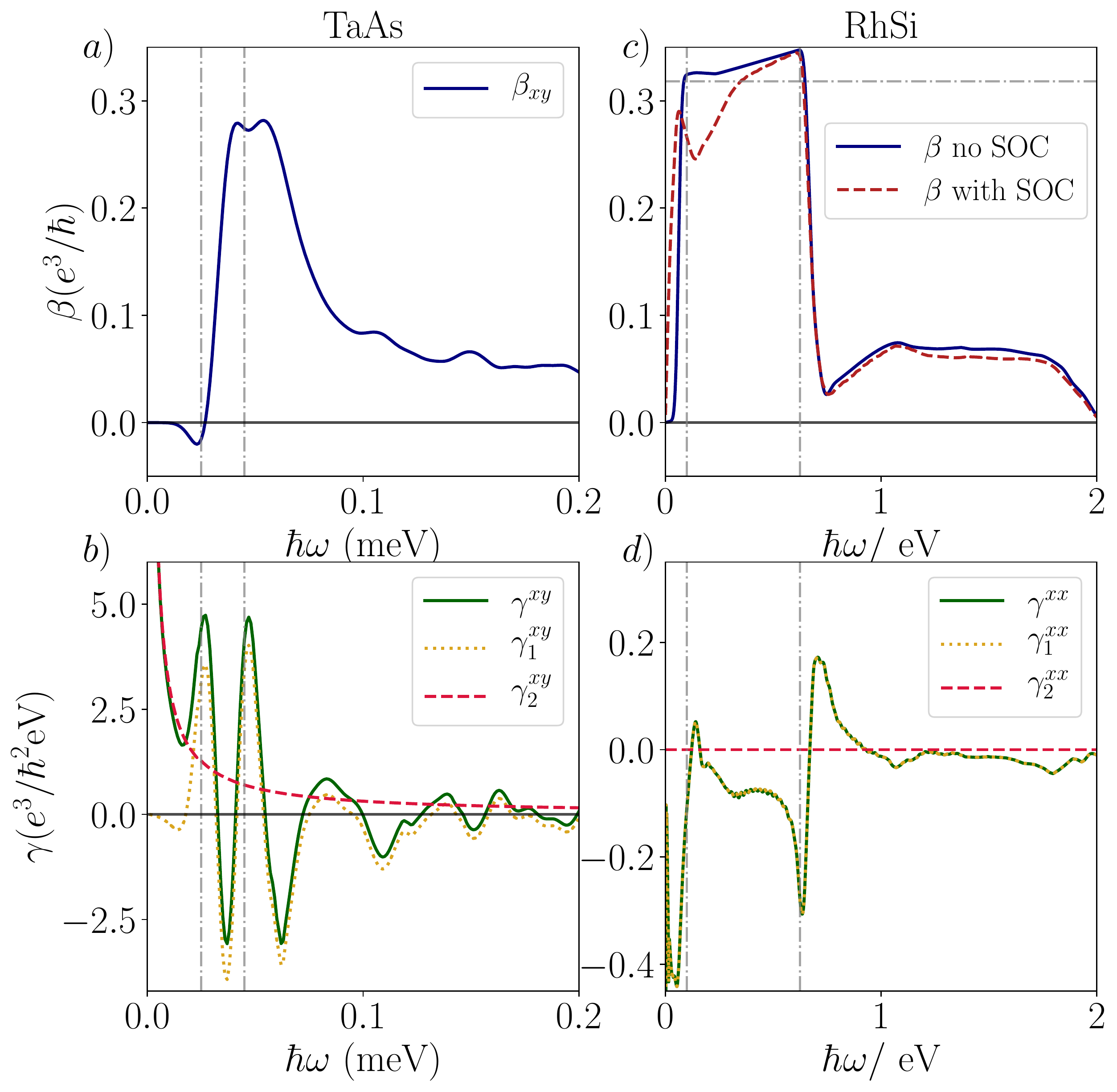}
\caption{First principles prediction for circular DFG. a) and b) show $\beta_{xy}$ and $\gamma_{xy}$ for TaAs with spin-orbit coupling (SOC). The activation frequencies of the W2 and W1 pairs of nodes at $25$ and $45$ meV (dashed vertical lines) mark the peaks in a) and b), in qualitative agreement with Fig.~\ref{fig1} b) and d). In c) and d) we show the trace $\beta$ (with and without SOC) and $\gamma_{xx}$ with SOC for RhSi. The activation frequencies of the threefold at $\Gamma$ and fourfold fermion at $R$, $0.1$ and $0.625$ eV respectively, are marked by dashed vertical lines. The horizontal dashed line in b) marks the effective monopole charge of 4 corresponding to $4\beta_0=0.318e^3/\hbar^2$. For b) and d) the total free-carrier contribution $\gamma$ (green-solid) is composed of the semiclassical ($\gamma_2$, red-dashed) and the novel free-carrier contribution ($\gamma_1$, orange-dotted), which qualitatively agree with Fig.~\ref{fig2} d).}
\label{fig3}
\end{figure}

In order to provide a qualitative prediction, we have calculated the different parts of the circular DFG for TaAs using density functional theory (DFT) (see~\cite{Supplementalphotoweyl} for details). The results are shown in Fig.~\ref{fig3} a) and b). 
We observe that $\gamma^{yx}$ shows the $1/\omega$ divergence, and several sign changes close to $2\mu_{W_i}$, where $\mu_{W_i}$ is the energy of the two types of Weyl nodes W1 and W2 present in TaAs~\cite{Weng2015,Huang:vn,Buckeridge:2016fb}. Consistently, $\beta^{yx}$ exhibits characteristic peaks around $2\mu_{W_{i}}$. These features follow qualitatively those expected for the traceless circular DFG components, shown in Fig.~\ref{fig3} b) and d).

Unlike TaAs, materials in the cubic space group 198, such as RhSi, CoSi or AlPt, are chiral and lack mirror symmetry. 
Taking RhSi as an example, and using the cubic nature of the space group we have calculated the diagonal components of the DFG using DFT (see Fig.~\ref{fig1} c) and d)). 
Consistent with Ref.~\cite{Flicker2018} we find that $\beta$ displays a flat region between the activation frequencies of the threefold and fourfold fermions. 
However the response is not exactly quantized: it is corrected by the presence of additional bands deviating from exact quantization even in the absence of spin-orbit coupling. 
These corrections are expected to decrease upon decreasing the chemical potential.
Lastly, the free-carrier contribution $\gamma_{xx}$ shows peaks and sign changes close to the activation frequencies of the threefold and fourfold fermions, consistent with Fig.~\ref{fig2} d).

\emph{Discussion - } In this work, we have described the circular DFG response in metals in the clean limit $\omega \gg \Delta \omega \gg \tau^{-1}$, where it is an intrinsic response of the band structure. We have found that it is composed of two contributions that oscillate out of phase, $\beta^{ab}$ and $\gamma^{ab}$: the trace of the former is quantized in chiral topological metals, and the latter is a free-carrier contribution that vanishes in the absence of a Fermi surface.

The measurement of the quantized trace of $\beta^{ab}$ can be achieved if the hierarchy of scales $\omega \gg \Delta \omega \gg \tau^{-1}$ is met. Interestingly evaluating if this condition is met is possible without prior knowledge of $\tau$: when $\Delta \omega \gg \tau^{-1}$ the interband circular DFG signal oscillates exactly out of phase with the incoming light, while in the dirty limit $\Delta \omega \ll \tau^{-1}$ it oscillates in phase. Similarly, when $\Delta \omega \gg \tau^{-1}$ the free-carrier circular DFG  $\gamma^{ab}$ oscillates in-phase and displays a characteristic singular $\omega$ dependence. The observation of an out-of-phase smooth component together with an in-phase component with singularities is therefore a strong signal that the clean limit has been achieved.

In the currently available RhSi, the plateau extends up to $\hbar\omega \sim 0.7$ eV~\cite{Rees19}, so setting $\Delta\omega = 100$ meV may allow scattering rates as large as $\tau^{-1} \sim 20 $ meV or $\tau \sim 200 $ fs. It should be noted that the quantized plateau is corrected in multifold materials by multi-band corrections~\cite{Flicker2018}, as confirmed by our DFT calculations, so it is desirable to search for new chiral Weyl semimetals with simpler band structures to reach exact quantization.

The relation of the free-carrier DFG response with the more often measured DC photogalvanic responses (where $\Delta \omega = 0$ with finite $\tau$) is quite  subtle. Both general arguments and explicit calculations have been used to argue that the free-carrier part of the DC photocurrent vanishes in the absence of a specific mechanism for dissipation~\cite{BelinicherTransient,BS80,BelinicherKinetic}. In the dirty limit ($\Delta \omega \ll \tau^{-1}$) there are $\tau$-independent disorder-induced contributions which can cancel the apparently intrinsic contributions~\cite{BelinicherTransient}, in a situation reminiscent of the anomalous Hall effect. In view of other disorder-induced photocurrent calculations~\cite{Konig17,Isobe18,Golub18,Sodemann19}, the extent to which this cancellation happens remains to be understood. Due to our assumption that $\Delta \omega \gg \tau^{-1}$ the above subtleties do not affect the DFG regime pertinent to this work. 

In conclusion, our work shows how difference frequency generation provides a disorder-independent route to separately measure photocurrent quantization, and a novel Fermi surface contribution beyond semiclassics. More broadly, it suggests that non-monochromatic responses encompass a rich set of experimental avenues to expose topology.

\emph{Acknowledgements} - We are grateful to L. Wu for enlightening discussions, and to L. Golub and M. Teixido for providing us with copies of Ref.~\cite{BelinicherTransient}. We acknowledge O. Matsyshyn and I. Sodemann for valuable discussions, and for sharing their manuscript on a study employing a similar formalism prior to publication \cite{intiack}. A.~G.~G.~acknowledges financial support from the Marie Curie programme under EC Grant agreement No.~653846. J.~E.~M. was supported by the Quantum Materials program at LBNL, funded by the US Department of Energy under Contract No. DE-AC02-05CH11231.


%

\newpage

\appendix

\begin{widetext}
\section{Appendix A: Difference frequency generation in the length gauge}\label{AppendixA}

The length gauge approach to the computation of non-linear response functions has been developed in detail for semiconductors in Refs.  \cite{GhahramaniSipe,AversaSipe,SipeShkrebtii,NastosSipe06,NastosSipe10}. The same approach can be used for metals in the non-interacting limit with a small number of modifications. In this formalism, the coupling of the bare Hamiltonian $H_0$ to the electric field takes the form $H = H_0 - e \vec E \cdot \vec r$ with $e<0$, and the current expectation value can be expressed in terms of the matrix elements of the density matrix $c_{nm} = \left<a^\dagger_n(k) a_m(k) \right>$ as the sum of interband and intraband pieces
\begin{align}
J(t) &= J_{\rm intra}(t) + \partial_t P_{\rm inter}(t), \\
P_{\rm inter}^a(t) &= e \int_k \sum_{nm} r^a_{nm}c_{mn}, \\
J^a_{\rm intra}(t) &= e \int_k \sum_{nm} D^a_{nm} c_{mn},
\end{align}
where $\int_k =\int d^3k/(2\pi)^3$ and $D^a_{nm} = v_{nn}^a \delta_{nm} - (e/\hbar) E^b(t) [r^b_{nm;a} +\delta_{nm}\epsilon^{dab}\Omega^d_n]$ is a generalized velocity which contains interband and intraband anomalous velocities, and $\Omega^a_n = \epsilon^{abc} \xi^c_{nn,b}$ is the Berry curvature. The intraband anomalous velocity, omitted in Ref. \cite{SipeShkrebtii} but included in Ref. \cite{AversaSipe}, is immaterial for insulators but important for metals, in particular because it gives rise to the semiclassical photogalvanic effects. 

The density matrix satisfies the dynamical equation $i\hbar \partial_t c = [H,c]$,  which for the matrix elements implies
\begin{equation}
\partial_t c_{mn} + i \omega_{mn}c_{mn}= \frac{eE^b(t)}{\hbar} \left[ - c_{mn;b}+ i \sum_p r^b_{mp}c_{pn}-c_{mp}r^b_{pn}\right],\label{Eq:dyn}
\end{equation}
To obtain the non-linear response functions, $c_{mn}$ are expanded iteratively to the desired order in the electric field  \cite{SipeShkrebtii}, starting with $c_{nm}^{(0)} = f_n \delta_{nm}$ where $f_n$ is the zero temperature Fermi function for band $n$. For insulators, $f_n=1$ for occupied bands and $f_n=0$ otherwise, but in the presence of a Fermi surface the $c_{nm}$ acquire extra pieces because $f_n$ depends on $k$. Expanding in the electric field {$E^b(t) = E^b_\beta e^{-i\omega_\beta t}$} where $\beta$ runs over different monochromatic components and is summed over the terms up to second order are 
\begin{align}
c_{mn}^{(1)}(t) &=  -\frac{e}{\hbar} E_\beta^b e^{-i\omega_\beta t} \left[
\mathcal{B}^b_{mn}
+i \frac{f_{m,b}\delta_{mn}}{\omega_\beta}\right], \label{firstE}\\
c_{mn}^{(2)}(t) &= \frac{e^2}{\hbar^2} E_\beta^b E_\gamma^c e^{-i\omega_\Sigma t} \left[ \frac{if_{nm}\left( \tfrac{r_{mn}^b}{\omega_\beta-\omega_{mn}}\right)_{;c} + \sum_p (r_{mp}^c \mathcal{B}_{pn}^b - \mathcal{B}_{mp}^b r_{pn}^c)}{\omega_\Sigma-\omega_{mn}} \right. \nonumber 
\\
&{\left.-\frac{f_{n,bc}\delta_{nm}}{\omega_\beta \omega_\Sigma}+ i \frac{r_{mn}^c f_{nm,b}}{\omega_\beta(\omega_\Sigma-\omega_{mn})}+i \frac{r_{mn}^b f_{nm,c}}{(\omega_\beta-\omega_{mn})(\omega_\Sigma-\omega_{mn})}\right]},
\label{secondE}
\end{align}
with $\mathcal{B}^b_{mn} = \frac{f_{nm}r_{mn}^b}{\omega_\beta-\omega_{mn}}$ and $\omega_\Sigma = \omega_\beta + \omega_\gamma$. The frequencies $\omega_\beta$ should be understood as $\omega_\beta + i \epsilon$ where the limit $\epsilon \rightarrow 0$ is taken after the thermodynamic limit. These coefficients were calculated in Ref.~\cite{Hipolito16}, yet we find some sign differences, e.g. the absence of a relative sign between the two terms in Eq.~\eqref{firstE}.

It is important to note that the diagonal ($n=m$) parts of these expressions, understood in the limit $\epsilon \rightarrow 0$, are valid only when the frequency denominators are strictly finite. In the absence of scattering mechanisms, taking the limits $\omega_\beta \rightarrow 0$ or $\omega_\Sigma \rightarrow  0$ will lead to divergences in $c_{nn}^{(1)}$ and $c_{nn}^{(2)}$ respectively. The dynamical equations for $c_{nn}$ are simply not defined in this limit.

These divergences are physical: they represent the response of an electron system in the idealized case of no relaxation. If we wish to know the response of the system when one of these frequencies is zero, scattering must be included rigorously in the calculation with the quantum kinetic equation~\cite{Culcer17}. As discussed previously~\cite{BelinicherTransient,BS80,BelinicherKinetic,Golub18}, this is of paramount importance in particular for the Fermi surface free-carrier contributions to the photocurrent, because certain disorder-induced mechanisms give rise to photocurrents that turn out to be independent of the disorder details. 
These contributions are not captured in the phenomenological approach that simply introduces a constant relaxation time in the dynamical equation. The same reasoning applies for the off diagonal $c_{nm}$ terms if the chermical potential is near a band degeneracy where $\omega_{nm}\rightarrow0$. 

In light of this discussion, the only intrinsic response that can be modeled without a scattering mechanism is the response in the limit $\omega_\Sigma\gg1/\tau$. This situation is familiar from the context of the anomalous Hall effect, which also has intrinsic and disorder induced contributions, and the intrinsic one can be measured as the AC Hall conductivity in the limit $\omega \rightarrow 0$ with $\omega \gg 1/\tau$. In the photocurrent context, the limit $\omega_\Sigma\gg1/\tau$ represents the frequency difference response to two slightly detuned monochromatic lasers. The response functions in this limit can also be used to compute the transient photocurrent response to pulsed light when the duration of the pulse is much shorter than $\tau$. 

With these caveats in mind, we now proceed to compute the response functions. As a warmup, we first review the linear response, where the presence of a Fermi surface gives to the Drude divergence. Collecting the terms to linear order, the full first order response is given by
\begin{align}
J^a(t) &= \partial_t P_{\rm inter}^a(t)+J^a_{\rm intra}(t)=\sum_{\omega_\beta=\pm \omega} \sigma^{ab}(\omega_\beta)E_\beta^b e^{-i\omega_\beta t},\\
\sigma^{ab}(\omega) &=\frac{e^2}{\hbar} \int_k \sum_{nm} \left[i \omega f_{nm}  \frac{r^a_{nm} r_{mn}^b}{\omega-\omega_{mn}}     - \delta_{nm}f_n \left(\frac{\omega_{n,ba}} {i\omega} + \epsilon^{abc} \Omega_c \right)\right]. \label{linearintra}
\end{align}
{
Where integration by parts and $v_{nn}^a = \omega_{n,a}$ were used. Note $\sigma^{ab}(\omega) = (\sigma^{ab}(-\omega))^*$. For further reference we can separate this into the absorptive (or dissipative) part $\sigma^{ab}_{\rm abs}(\omega)$ and reactive (or non-dissipative) part $\sigma^{ab}_{\rm rea}(\omega)$ as
\begin{align}
\sigma^{ab}_{\rm abs} = \frac{1}{2}\left[\sigma^{ab} + (\sigma^{ba})^*\right]  &=\frac{e^2\pi}{\hbar} \int_k \sum_{nm} \left[ -\omega f_{nm}  r^a_{nm} r_{mn}^b \delta(\omega-\omega_{mn})- \delta_{nm}f_n \omega_{n,ba} \delta(\omega) \right]\\
\sigma^{ab}_{\rm rea} = \frac{1}{2}\left[\sigma^{ab} - (\sigma^{ba})^*\right] &=\frac{e^2}{\hbar} \int_k \sum_{nm} \left[i \omega f_{nm} \mathcal{P} \frac{r^a_{nm} r_{mn}^b}{\omega-\omega_{mn}} - \delta_{nm}f_n \left(\mathcal{P}\frac{\omega_{n,ba}} {i\omega} + \epsilon^{abc} \Omega_c \right)\right].\label{linearreactive}
\end{align}
Note both parts have real and imaginary parts. In the presence of time reversal symmetry, $\sigma^{ab}_{\rm abs}$ is real, $\sigma^{ab}_{\rm rea}$ is imaginary, and both are symmetric under $a \leftrightarrow b$.} The equivalent expression in the velocity gauge and how to transform it into this one are discussed in \cite{GhahramaniSipe}, see Eq. 1.9.

Next we consider the second order contributions to the current, which take the general form 
\begin{align}
\partial_t P_{\rm inter}^a(t) &= -i\omega_\Sigma\chi^{abc,(2)}_{inter}(\omega_\beta,\omega_\gamma)E_\beta^b E_\gamma^c e^{-i\omega_\Sigma t}, \\
J^a_{\rm intra}(t) &= \chi^{abc,(2)}_{intra}(\omega_\beta,\omega_\gamma) E_\beta^b E_\gamma^c e^{-i\omega_\Sigma t}. \label{Jintra_gen}
\end{align}
$\chi^{abc}_{\rm inter}$ has the insulator contribution derived in Ref. \cite{SipeShkrebtii} plus those obtained from the terms containing $f_{n,a}$ in Eq. \eqref{secondE}. Due to the extra factor $\omega_\Sigma$, both contributions to $\partial_t P_{\rm inter}$, which give rise to an optical rectification current, are subdominant when $\omega_\Sigma$ is small and will be neglected compared to those in $J_{\rm intra}^a$ in this work.  

The dominant contributions to frequency difference generation come from $\chi^{abc}_{\rm intra}$, which is given by 

\begin{equation}
\chi^{abc}_{\rm intra}(\omega_\beta,\omega_\gamma) =\frac{1}{-i\omega_\Sigma} \eta^{abc}(\omega_\beta,\omega_\gamma) + \sigma^{abc}(\omega_\beta,\omega_\gamma)+\rho_2^{abc}(\omega_\beta,\omega_\gamma). \label{chiintra2}
\end{equation}
The first two terms in this expression are the standard ones obtained for semiconductors~\cite{SipeShkrebtii}. They can be expressed in terms of single frequency tensors  
\begin{align}
\eta^{abc}(\omega_\beta,\omega_\gamma) &= \Gamma^{abc}(\omega_\beta) + \Gamma^{acb}(\omega_\gamma)  \label {inj},\\
\sigma^{abc}(\omega_\beta,\omega_\gamma) &=  \Lambda^{abc}(\omega_\beta) + \Lambda^{acb}(\omega_\gamma),
\end{align}
which take the form 
\begin{align}
\Gamma^{abc}(\omega)&= i\frac{C}{2} \int_k \sum_{nm} f_{nm} \Delta^a_{nm} \frac{r^c_{nm}r^b_{mn}}{\omega_{mn}-\omega},  \\
\Lambda^{abc}(\omega) &= -\frac{C}{2} \int_k \sum_{nm} f_{nm} \frac{r^c_{nm;a}r^b_{mn}}{\omega_{mn}-\omega},\label{Lambda}\end{align}
where $C=e^3/\hbar^2$. Using $(r^c_{nm})^*=r^c_{mn}$, $(r^c_{nm;a})^*=r^c_{mn;a}$, we see these single frequency tensors satisfy $\Gamma^{abc}(-\omega) = \left[\Gamma^{abc}(\omega)\right]^*$ and $\Lambda^{abc}(-\omega) =  \left[\Lambda^{abc}(\omega)\right]^*$ so both tensors satisfy the standard Kramers-Kronig relations. 

Finally, there is also the new Fermi surface contribution 
\begin{align}
\rho^{abc}_2(\omega_\beta,\omega_\gamma) &= -\frac{C}{2} \int_k \sum_{n} \left[ i\Omega^d_n\left(\frac{\epsilon^{dac} f_{n,b}}{\omega_\beta}+\frac{\epsilon^{dab} f_{n,c}}{\omega_\gamma}\right)+\frac{v_{nn}^af_{n,bc}}{\omega_\beta \omega_\gamma} \right].    
\end{align}
In the last term we have used $\frac{1}{\omega_\beta} + \frac{1}{\omega_\gamma} = \frac{\omega_\Sigma}{\omega_\beta \omega_\gamma}$. 
$\rho^{abc}_2$ is new compared to the insulating case and contains two terms: the first reproduces the the semiclassical contributions \cite{Deyo09,MooreOrenstein,sodemannfu,Morimoto16,IHU16,Rostami17,ZSY18} originating in the diagonal anomalous velocity. The second term is purely classical \cite{sodemannfu}. 

\subsection{Frequency difference generation at small $\Delta \omega$}

We now consider the second order response to two monochromatic beams of frequencies $\omega_1$ and $\omega_2$, $E^b(t) = (E^b_1 e^{-i\omega_1 t}+E^{b*}_1 e^{i\omega_1 t} + E^b_2 e^{-i\omega_2 t}+E^{b*}_2 e^{i\omega_2 t})$. Our aim is to take slightly detuned frequencies $\omega_1 =\omega- \tfrac{\Delta\omega}{2}$ and $\omega_2 =\omega + \tfrac{\Delta\omega}{2}$ and look at the frequency difference generation terms which oscillate with a low frequency $\Delta\omega =\omega_2-\omega_1$ (i.e. when $\omega_\Sigma = \pm \Delta\omega$). Keeping only these terms in the sum in Eq. \eqref{Jintra_gen} we have

\begin{align}
J^a_{\rm intra}(t) 
&= 2\chi^{abc,(2)}_{intra}(\omega- \tfrac{\Delta \omega}{2},-\omega- \tfrac{\Delta \omega}{2}) e^{-i\Delta \omega t} E_1^b E_2^{c*} +
2\chi^{abc,(2)}_{intra}(\omega+ \tfrac{\Delta \omega}{2},-\omega+ \tfrac{\Delta \omega}{2}) e^{i\Delta \omega t} E_2^b E_1^{c*}\nonumber\\
&=2 \sum_{\omega_\Sigma = \pm \Delta\omega}  \chi^{abc,(2)}_{intra}(\omega- \tfrac{\omega_\Sigma}{2},-\omega- \tfrac{\omega_\Sigma}{2}) e^{-i\omega_\Sigma t} E^b E^{c*}, \label{JDFG}
\end{align}
where we have used the symmetry of $\chi^{abc,(2)}(\omega_\beta,\omega_\gamma) = \chi^{acb,(2)}(\omega_\gamma,\omega_\beta)$ which originates from the exchange of dummy indices, and in the last step we have considered the case where the two beams have the same polarization $E_1^b = E_2^b$ for simplicity.

We now consider the limit where $|\omega_\Sigma| \ll \omega$. The dominant term in this limit comes from the $1/\omega_\Sigma$ pole accompanying $\eta^{abc}$ in Eq. \eqref{chiintra2}. Since we are also interested in the next leading order terms which are independent of $\omega_\Sigma$, we need to expand $\eta^{abc}$ to first order in $\omega_\Sigma$.  We therefore have 
\begin{align}
\chi^{abc,(2)}_{intra}(\omega- \tfrac{\omega_\Sigma}{2},-\omega- \tfrac{\omega_\Sigma}{2}) &= \frac{1}{-i\omega_\Sigma}\left( \eta^{abc}(\omega,-\omega) + \omega_\Sigma \partial_{\omega_\Sigma}\eta^{abc}(\omega- \tfrac{\omega_\Sigma}{2},-\omega- \tfrac{\omega_\Sigma}{2})|_{\omega_\Sigma=0} \right) \nonumber\\
&+ \sigma^{abc}(\omega,-\omega)+\rho_2^{abc}(\omega,-\omega) +O(\omega_\Sigma). \label{chi_intra_1}
\end{align}
When evaluating these terms explicitly,
they will naturally separate into resonant (or absorptive) terms, proportional to delta functions, and non-resonant (or reactive) terms proportional to principal value integrals. In the absence of time reversal symmetry, both resonant and non-resonant parts are in general complex. Using that $\Gamma^{acb}(-\omega) = \left[\Gamma^{acb}(\omega)\right]^*$ and $\Lambda^{acb}(-\omega) = \left[\Lambda^{acb}(\omega)\right]^*$, the four terms in Eq. \eqref{chi_intra_1} can be computed as 
\begin{align}
\eta^{abc}(\omega,-\omega) &= \Gamma^{abc}(\omega) + [\Gamma^{acb}(\omega)]^* = \Gamma^{abc}_{\rm abs}(\omega), \\
\partial_{\omega_\Sigma} \eta^{abc}(\omega- \tfrac{\omega_\Sigma}{2},-\omega- \tfrac{\omega_\Sigma}{2})|_{\omega_\Sigma=0} &=
\partial_{\omega_\Sigma}[\Gamma^{abc}(\omega- \tfrac{\omega_\Sigma}{2})+   \Gamma^{acb}(-\omega- \tfrac{\omega_\Sigma}{2})] 
 = -\frac{\partial_{\omega}}{2} \Gamma_{\rm rea}^{abc}(\omega), \label{d_eta} \\
 \sigma^{abc}(\omega,-\omega) &= \Lambda^{abc}(\omega) + [\Lambda^{acb}(\omega)]^* = \Lambda_{\rm abs}^{abc}(\omega)+\Lambda_{\rm rea}^{abc}(\omega) \label{sigma_ph}, \\
 \rho_2^{abc}(\omega,-\omega) &= 
-\frac{C}{2\omega} \int_k \sum_{n}f_{n,a} \left[ i\Omega^d_n\epsilon^{dbc} +
\mathcal{P}\frac{\omega_{n,bc}}{\omega} \right],\label{rho22}
\end{align}
where the different absorptive and reactive parts take the form
\begin{align}
\Gamma^{abc}_{\rm abs}(\omega) &= {\color{blue}-} \pi C \int_k \sum_{nm} f_{nm} \Delta^a_{nm} r^c_{nm}r^b_{mn}\delta(\omega_{mn}-\omega),  \\
\Gamma^{abc}_{\rm rea}(\omega) &= iC \int_k \sum_{nm} f_{nm} \Delta^a_{nm} r^c_{nm}r^b_{mn}\mathcal{P}\frac{1}{\omega_{mn}-\omega},\\
\Lambda_{\rm abs}^{abc}(\omega) &\equiv-\frac{iC \pi}{2} \int_k \sum_{nm} f_{nm}(r^c_{nm;a}r^b_{mn}-r^c_{nm}r^b_{mn;a}) \delta(\omega_{nm}-\omega), \\
\Lambda_{\rm rea}^{abc}(\omega) &\equiv-\frac{C}{2} \int_k \sum_{nm} f_{nm} \mathcal{P} \frac{(r^c_{nm}r^b_{mn})_{,a}}{\omega_{mn}-\omega}. \label{LambdaNR}
\end{align}

Note the covariant derivatives in Eq. \eqref{LambdaNR} turned into regular ones because the connections drop out. Also note that by construction $\Lambda_{\rm abs}^{abc}$ and $\Lambda^{abc}_{\rm rea}$ denote only the resonant and non-resonant parts of Eq. \eqref{Lambda} which give rise to a real current, which are either real and symmetric or imaginary and antisymmetric. Also note in Eq. \eqref{rho22} we used $\epsilon^{abc}\Omega^c = \xi_{a,b}-\xi_{b,a}$, in last term we  used $v^a_{nn} = \omega_{n,a}$, and we integrated by parts in both terms. 

Putting all terms together we have 
\begin{align}
\chi^{abc,(2)}_{intra}(\omega- \tfrac{\omega_\Sigma}{2},-\omega- \tfrac{\omega_\Sigma}{2}) &= \frac{\Gamma_{\rm abs}^{abc}(\omega) - \omega_\Sigma\frac{\partial_{\omega}}{2} \Gamma_{\rm rea}^{abc}(\omega)}{-i\omega_\Sigma} + \Lambda_{\rm abs}^{abc}(\omega) +\Lambda_{\rm rea}^{abc}(\omega)+ \rho_2^{abc}(\omega). \label{chi_intermediate}
\end{align}
While this expression appears to contain reactive terms, we next show that these vanish. For this we follow \cite{NastosSipe10} and use  $\partial_\omega \left(\tfrac{1}{\omega_{nm}-\omega}\right) = -\partial_{\omega_{mn}} \left(\tfrac{1}{\omega_{nm}-\omega}\right) = - \frac{1}{\Delta^a_{mn}} \partial_{k_a}\left(\tfrac{1}{\omega_{nm}-\omega}\right)$ and integrate by parts in Eq. \eqref{d_eta} to show that 
\begin{align}
-\frac{\partial_{\omega}}{2} \Gamma_{\rm rea}^{abc}(\omega) &=   -i\frac{C}{2} \int_k \sum_{nm}\mathcal{P} \frac{f_{nm,a} r^c_{nm}r^b_{mn} +f_{nm} (r^c_{nm}r^b_{mn})_{,a}}{\omega_{mn}-\omega}  = -i\rho_1^{abc}(\omega) + i\Lambda_{\rm rea}^{abc}(\omega), \label{part}
\end{align}
where the first term is a new Fermi surface contribution
\begin{align}
\rho_1^{abc}(\omega) &= \dfrac{C}{2} \int_k \sum_{nm} f_{nm,a}\mathcal{P} \frac{r^c_{nm}r^b_{mn}}{\omega_{mn}-\omega}.\label{rho1}
\end{align}
Finally, substituting Eq. \eqref{part} into Eq. \eqref{chi_intermediate} we get
\begin{align}
\chi^{abc,(2)}_{intra}(\omega- \tfrac{\omega_\Sigma}{2},-\omega- \tfrac{\omega_\Sigma}{2}) &= 
\frac{\Gamma_{\rm abs}^{abc}(\omega)}{-i\omega_\Sigma} + \Lambda_{\rm abs}^{abc}(\omega) + \rho_1^{abc}(\omega) + \rho_2^{abc}(\omega).
\end{align}
Indeed this response contains no reactive part (even in the absence of time-reversal symmetry). The final expression for the time dependent DFG current \eqref{JDFG} takes the form 
\begin{align}
J^a_{\rm intra}(t) = \left[ \Gamma^{abc}_{\rm abs}(\omega) \frac{\sin \Delta\omega t}{\Delta\omega} +\left(\Lambda_{\rm abs}^{abc}(\omega) + \rho^{abc}(\omega) \right) \cos \Delta\omega t \right] 4 E^b E^{c*},
\label{eq:reftotalcurrentapp}
\end{align}
where we defined the total free-carrier contribution $\rho^{abc}(\omega)=\rho_1^{abc}(\omega)+\rho_2^{abc}(\omega)$. In the main text we drop the ``intra" label since $J_{\rm intra}$ is the leading contribution to the total current in the limit of small $\Delta\omega$. 

It is worth noting that the total free-carrier contribution takes the form 
\begin{align}
\rho^{abc}(\omega) =  \dfrac{C}{2\omega} \int_k \sum_{nm}\left[  f_{nm,a}\omega \mathcal{P} \frac{r^c_{nm}r^b_{mn}}{\omega_{mn}-\omega}-\delta_{nm} f_{n,a} \left[ i\Omega^d_n\epsilon^{dbc} +
\mathcal{P}\frac{\omega_{n,bc}}{\omega} \right]\right],
\end{align}
which, remarkably, is obtained from the reactive linear conductivity $\sigma^{ab}_{\rm rea}$ in Eq.   \eqref{linearreactive} making the replacement $f_n \rightarrow f_{n,a}/\omega$. Therefore, for every contribution to the reactive conductivity there is an intraband photocurrent. And because of the extra $k$ derivative in $f_{n,a}$, the time-reversal even parts of the photocurrent correspond to the time reversal odd parts of the reactive linear conductivity and viceversa. In the presence of time-reversal invariance, $\sigma^{ab}_{\rm rea} = \sigma^{ab}_{\rm Hall}$ is the optical Hall conductivity, and only two terms survive
\begin{align}
\rho^{abc}(\omega) =  \dfrac{iC}{2\omega} \int_k \sum_{nm}\left[  f_{nm,a}\omega \mathcal{P} \frac{{\rm Im}[r^c_{nm}r^b_{mn}]}{\omega_{mn}-\omega}-\delta_{nm} f_{n,a}  \Omega^d_n\epsilon^{dbc} \right],
\end{align}
which will lead to the contributions discussed in the main text, $\gamma^{ab}_{1}$ and $\gamma^{ab}_{2}$, when transformed into two index tensors as discussed below. 

\section{Kramers-Kronig relations}
Here we show the explicit Kramers-Kronig relations. For any complex function which is analytic in the upper half plane $\chi(\omega)$ and satisfies $\chi(-\omega) = \chi(\omega)^*$ we have the usual relations
\begin{align}
{\rm Re}\chi(\omega) &= \frac{2}{\pi} \mathcal{P}\int_0^\infty d\omega' \frac{\omega' {\rm Im}\chi(\omega')}{\omega'^2-\omega^2}\\
{\rm Im}\chi(\omega) &= -\frac{2\omega}{\pi} \mathcal{P}\int_0^\infty d\omega' \frac{{\rm Re}\chi(\omega')}{\omega'^2-\omega^2}
\end{align}
Recalling that $\sigma^{ab}(-\omega) = [\sigma^{ab}(\omega)]^*$, $\Gamma^{abc}(-\omega) = [\Gamma^{abc}(\omega)]^*$ and $\Lambda^{abc}(-\omega) = [\Lambda^{abc}(\omega)]^*$, the Kramers-Kronig relations can be used for any of these three tensors directly. However, it should be noted that in the absence of time-reversal symmetry real and imaginary parts do not map to absorptive and reactive parts. Since one is usually interested in computing the absorptive parts numerically because integrating delta functions is stable, and then reproducing the reactive parts with Kramers-Kronig, it is useful to spell out these relations solving for the reactive components. For the non-linear response tensors $\Gamma^{abc}$ and $\Lambda^{abc}$ we have
\begin{align}
{\rm Re}[\Gamma^{abc}_{\rm rea}(\omega)] &= \frac{2}{\pi} \mathcal{P}\int_0^\infty d\omega' \frac{\omega' {\rm Im}[\Gamma^{abc}_{\rm abs}(\omega')]}{\omega'^2-\omega^2},\\
{\rm Im}[\Lambda^{abc}_{\rm rea}(\omega)] &= -\frac{2\omega}{\pi} \mathcal{P}\int_0^\infty d\omega' \frac{{\rm Re}[\Lambda^{abc}_{\rm abs}(\omega')]}{\omega'^2-\omega^2},    \\
{\rm Im}[\Gamma^{abc}_{\rm rea}(\omega)] &=- \frac{2\omega}{\pi} \mathcal{P}\int_0^\infty d\omega' \frac{{\rm Re}[\Gamma^{abc}_{\rm abs}(\omega')]}{\omega'^2-\omega^2},\\
{\rm Re}[\Lambda^{abc}_{\rm rea}(\omega)] &= \frac{2}{\pi} \mathcal{P}\int_0^\infty d\omega' \frac{\omega'{\rm Im}[\Lambda^{abc}_{\rm abs}(\omega')]}{\omega'^2-\omega^2}.   
\end{align}

One application of these identities is the numerical computation of $\rho_1^{abc}$, since its definition in Eq. \eqref{rho1} might be unstable due to the presence of the energy denominator. Solving for $\rho_1^{abc}$ in Eq. \eqref{part} we also find 
\begin{align}
\rho_1^{abc}(\omega) =-i\frac{\partial_{\omega}}{2} \Gamma_{\rm rea}^{abc}(\omega) +\Lambda_{\rm rea}^{abc}(\omega), \label{rho1KK}
\end{align}
which allows to compute $\rho_1^{abc}$ via Kramers-Kronig relations. We have found this to be more stable for ab-initio calculations.

\subsection{Explicit real and imaginary parts}

Following Ref.~\cite{SipeShkrebtii}, all these results can be rewritten in a way that naturally separates real vs imaginary and symmetric vs antisymmetric parts. To do so, we note that the sum over states counts every pair twice,  $\sum_{nm} = \sum_{n>m} +  \sum_{n<m}$ (the terms with $n=m$ are excluded since $f_{nn}=0$). One may rewrite all expressions with sums of the type $\sum_{n>m}$ only by relabeling dummy indices, and then interchanging them, and using that $f_{nm} = -f_{mn}$, $\omega_{nm} = - \omega_{mn}$, $r^a_{nm}(k) = (r_{mn}^a(k))^*$. The linear response conductivity reads
\begin{align}
\sigma^{ab}(\omega) =\frac{e^2}{\hbar} \int_k \left[(-i \omega) \sum_{n>m}f_{nm}\left( {\rm Re}[r^a_{nm} r_{mn}^b]F_+ + i{\rm Im}[r^a_{nm} r_{mn}^b]F_- \right)     -\sum_{n} \left( \frac{v_{nn}^a f_{n,b}}{-i\omega} + f_n \epsilon^{abc} \Omega_c \right)\right],\label{linearintegrand}
\end{align}
where we define the sum and difference of energy denominators 
\begin{align}
F_{\pm}(\omega_{mn},\omega) &= \frac{1}{\omega_{mn}-\omega-i\epsilon} \pm \frac{1}{\omega_{mn}+\omega+i\epsilon},
\end{align}
with the $i\epsilon$ factors made explicit. These functions should be
interpreted as $F_{\pm} = \frac{1}{\omega_{mn}-\omega} \pm \frac{1}{\omega_{mn}+\omega} + i\pi [-\delta(\omega_{mn}-\omega)\pm\delta(\omega_{mn}+\omega)]$ and satisfy
\begin{align}
F_{\pm}(\omega_{mn}, -\omega) = \pm  (F_{\pm}(\omega_{mn}, \omega))^*\label{conjF}.
\end{align}
For the non-linear response we find 
\begin{align}
\label{eq:GammaR}
\Gamma_{\rm abs}^{abc}(\omega) &=\dfrac{C}{2} \int_k \sum_{n>m}f_{nm}\Delta^a_{mn}\left[ i {\rm Im}(r^c_{nm}r^b_{mn}) {\rm Im}F_+ + {\rm Re}(r^c_{nm}r^b_{mn}) {\rm Im}F_-\right],\\
\label{eq:LambdaR}
\Lambda_{\rm abs}^{abc}(\omega) &=-\dfrac{C}{2} \int_k \sum_{n>m}f_{nm}\left[{\rm Im}(r^c_{nm;a}r^b_{mn}+r^b_{nm;a}r^c_{mn}) {\rm Im}F_- + i{\rm Re}(r^c_{nm;a}r^b_{mn} - r^b_{nm;a}r^c_{mn}){\rm Im}F_+\right],\\
\label{eq:rho1}
\rho_1^{abc}(\omega) &= \dfrac{C}{2} \int_k \sum_{n>m} f_{nm,a} \left(i{\rm Im}[r^c_{nm} r_{mn}^b] {\rm Re}F_- + {\rm Re}[r^c_{nm} r_{mn}^b] {\rm Re}F_+  \right), \\
\label{eq:rho2}
\rho_2^{abc}(\omega) &= 
-\frac{C}{2\omega} \int_k \sum_{n}f_{n,a} \left[ i\Omega^d_n\epsilon^{dbc} +
\mathcal{P}\frac{\omega_{n,bc}}{\omega} \right].
\end{align}
Recalling that $(r^c_{nm}r^b_{mn})^* = r^b_{nm}r^c_{mn}$ we see that each of these four terms contains a symmetric, real part which contributes to the linear photogalvanic effect (LGPE) plus an antisymmetric imaginary part which contributes to the circular photogalvanic effect (CPGE).  The contributions in $\Gamma_{\rm abs}^{abc}(\omega)$ give rise to an injection current (which grows linearly in time) while those in $\Lambda_{\rm abs}^{abc}(\omega)$ and $\rho_{1,2}^{abc}(\omega)$ give rise to a current that is constant in time. 

Finally, we can now check how the different pieces transform under the time reversal operation. This symmetry imposes $\omega_{nm}(k) = \omega_{nm}(-k)$, $r^a_{nm}(-k) = r_{mn}^a(k)$, $\Delta^a_{nm}(-k) = -\Delta^a_{nm}(k)$ and $r_{nm;a}^c(k) = -r_{mn;a}^c(-k)$. The last one follows since the operator $D_a = \partial_{k_a} + i(\xi^a_{nn}-\xi^a_{mm})$ gives an extra minus sign. Splitting integrals into $\int_k F(k) = 1/2 [\int_k F(k) + \int_{-k} F(-k)] =1/2 [\int_k (F(k) + F(-k))] $. we realize that 
\begin{align}
{\rm Re}\Gamma_{\rm abs}^{abc}(\omega) =  {\rm Im}\Lambda_{\rm abs}^{abc}(\omega) = {\rm Re}\rho_1^{abc}(\omega) ={\rm Re}\rho_2^{abc}(\omega) =0,
\end{align}
while the rest of contributions remain the same. With time-reversal symmetry, the injection current and Fermi surface contributions are therefore purely circular, while the current coming from $\Lambda_{\rm abs}^{abc}(\omega)$ is purely linear and is identified as the shift current. 

Also note that in the limit of  $\omega=0$, ${\rm Re}F_- = 0$ but ${\rm Re}F_+ = 1/\omega_{nm}$ is finite, so this term gives an additional semiclassical contribution (beyond the Berry curvature dipole and the Drude peak derivative). This contribution was found in Ref.~\cite{Gao14} by deriving the semiclassical equations of motion to second order in the electric field. 

\section{Mapping to two index tensors}

To connect to Eq.~\eqref{eq:allDFG} in the main text we rewrite Eq.~\eqref{eq:reftotalcurrentapp} in terms of two rank tensors:
\begin{align}
J^a_{\rm intra}(t) = 4\left[ \beta^{ab}(\omega) \frac{\sin \Delta\omega t}{\Delta\omega} +\left(\gamma^{ab}_1 (\omega) + \gamma_2^{ab}(\omega)\right) \cos \Delta\omega t \right] \epsilon^{blm}E_l E^{*}_m + 2\cos(\Delta \omega t) \; \sigma^{abc}(\omega)  (E^bE^{c*} +E^cE^{b*}),
\label{eq:reftotalcurrentapptwoindex}
\end{align}
where we used that in the presence of time reversal symmetry $\Gamma^{abc}_{\rm abs}, \rho_1^{abc}$ and $\rho_2^{abc}$ are antisymmetric and can be written in terms of a two index tensor, such that $\beta^{ad} = \tfrac{1}{2} \epsilon^{dbc} \Gamma_{\rm abs}^{abc}$, $\gamma^{ad}_1 = \tfrac{1}{2} \epsilon^{dbc} \rho_1^{abc}$ and $\gamma^{ad}_2 = \tfrac{1}{2} \epsilon^{dbc} \rho_2^{abc}$.
Using the definitions Eqs.~(\ref{eq:GammaR}) to \eqref{eq:rho2}, assuming $\omega>0$, and using that ${\rm Re}[F_-(\omega_{mn},\omega)] = 2\omega/(\omega_{nm}^2-\omega^2)$ and that ${\rm Im }[F_+(\omega_{mn},\omega)] = \pi [-\delta(\omega_{mn}-\omega)+\delta(\omega_{mn}+\omega)]$ we arrive at the equations in the main text which we repeat here for completeness
\begin{align}
{\beta^{ab}}(\omega) &= {i\frac{\pi C}{4}} \int_k {\sum_{n>m}} f_{nm} \Delta^a_{mn}{\rm Im} [r^d_{nm}r^c_{mn}] \epsilon^{bcd} \delta(\omega_{mn}-\omega), \label{eq:injectionapp}\\
{\sigma^{abc}}(\omega) &= \frac{\pi C}{2} \int_k {\sum_{n>m}} f_{nm} {\rm Im} [r_{mn}^b r_{nm;a}^c]  \delta(\omega_{mn}-\omega),\label{lambdaapp} \\
\gamma_1^{ab}(\omega) &= \frac{iC}{2} \int_k {\sum_{n>m}} \frac{\omega f_{nm,a}}{\omega_{nm}^2-\omega^2}  {\rm Im}[\epsilon^{bcd}r^d_{nm}r^c_{mn}],\label{FSintrabandapp}\\
\gamma_2^{ab} (\omega)&= -\frac{iC}{2\omega} \int_k  \sum_n  \left[(f_{n,a} \Omega_n^b - \delta^{ab} f_{n,c}\Omega_n^c)\right].\label{eq:semiclapp}
\end{align}
The analog of Eq. \eqref{rho1KK} for the two-index tensor $\gamma_1^{ab}$ assuming time-reversal symmetry is  
\begin{align}
\gamma_1^{ab}   =-i\frac{\partial_{\omega}}{2} \beta_{\rm rea}^{ab}(\omega) +\tfrac{i}{2} \epsilon^{acd}{\rm Im}[\Lambda_{\rm rea}^{bcd}(\omega)],\label{KKtouse}
\end{align}
where  
\begin{align}
\beta_{\rm rea}^{ab}(\omega) =  i \frac{C}{2} \int_k \sum_{nm} f_{nm} \Delta^a_{nm} \epsilon^{bcd}r^d_{nm}r^c_{mn} \mathcal{P}\frac{1}{\omega_{mn}-\omega}.
\end{align}

\section{Appendix B: Calculation for tilted Weyl node}

Here we present the computation of the circular photogalvanic tensors $\beta^{ab}$, $\gamma_1^{ab}$, $\gamma_2^{ab}$ for a general two band model of the form
\begin{equation}
H = \sigma_i v_{ij} k_j + u_i k_i -\mu,
\end{equation}
where $v_{ij}$ is a general Fermi velocity matrix that can always be made symmetric by an appropriate choice of the $\sigma_i$, $u_i$ is the tilt vector and $\mu>0$ is the chemical potential. Some of our results generalize those presented in Refs.\cite{Chan16,Ran17,Carbotte16Tilt,Steiner17}.

We work in the coordinate system given by crystallographic axes, where both $v_{ij}$ and $u_i$ will in general be arbitrary, as for a Weyl node at a generic point in the Brillouin zone there are no symmetry constraints on them. Symmetries will constrain the total tensors after adding the contributions of all nodes related by symmetry. This model has energies $\omega_1 = \vec u \vec k - \sqrt{v_{ij}k_j v_{il}k_l}$ and $\omega_2 = \vec u \vec k + \sqrt{v_{ij}k_j v_{il}k_l}$ and the Fermi functions are $f_1 = 1$ and $f_2 = \Theta(\mu - \omega_2)$. 

The expressions for the CPGE tensors in the two-band, clean limit read
\begin{align}
\beta^{ab} &= \frac{i \pi C}{4} \int_k f_{12} \Delta^a_{12} \Omega_2^b \delta(\omega_{21}-\omega), \\
\gamma_1^{ab} &= \frac{iC}{2} \int_k (-f_{2,a}) \Omega_1^b \frac{\omega}{\omega_{12}^2-\omega^2},\\
\gamma_2^{ab} &= \frac{-iC}{2\omega} \int_k (f_{2,a} \Omega_2^b - \delta^{ab} f_{2,c}\Omega_2^c).
\end{align}
The calculation can be simplified by defining a new coordinate $k'_i = v_{ij}k_j$, in terms of which we have
\begin{align}
\partial_{k_a} \omega_\alpha &= v_{ab} \partial_{k'_b} \omega_\alpha, \\
\Omega^a(k) &= \epsilon^{abc} v_{bm} v_{cn} \epsilon^{lmn} \frac{k'^l}{4 k'^3} = {\rm det} v \; (v^{-1})_{la} \frac{k'^l}{2 k'^3}.
\end{align}
Using these relations, we can perform the change of variables in the integral with $\int_k = \tfrac{1}{|{\rm det}v|} \int_{k^{'}}$, obtaining 
\begin{align}
\beta^{ab} &= \chi v^{ac}(v^{-1})^{bd} \beta'^{cd}, \\
\gamma_1^{ab} &= \chi v^{ac}(v^{-1})^{bd} \gamma_1'^{cd}, \\
\gamma_2^{ab} &= \chi v^{ac}(v^{-1})^{bd} \gamma_2'^{cd}, 
\end{align}
where $\chi = \tfrac{{\rm det}v}{|{\rm det}v|} = \pm 1$ is the chirality of the node and 
\begin{align}
\beta'^{ab} &= \frac{i \pi C}{4} \int_{k'} f_{12} \Delta^a_{12} \Omega_2^b \delta(\omega_{21}-\omega), \\
\gamma_1'^{ab} &= \frac{iC}{2} \int_{k'} (-f_{2,a}) \Omega_1^b \frac{\omega}{\omega_{12}^2-\omega^2},\\
\gamma_2'^{ab} &= \frac{-iC}{2\omega}\int_{k'} (f_{2,a} \Omega_2^b - \delta^{ab} f_{2,c}\Omega_2^c),
\end{align}
where all quantities are now computed from the Hamiltonian $H = \sigma_i k_i' + \tilde{u}_i k'_i$, with $\tilde{u}_i = (v^{-1})_{ij} u_j$. 

The tensor structure of $\beta'^{ij}$, $\gamma_1'^{ij}$ and $\gamma_2'^{ij}$ can only contain $\tilde{u}_i\tilde{u}_j/\tilde{u}^2$ and $\delta^{ij}$, and by separating into trace and traceless prefactors we have
\begin{align}
\beta'^{ij}=  \frac{\delta^{ij}}{3}\beta+\left[\frac{\delta^{ij}}{3}-\frac{\tilde{u}_i\tilde{u}_j}{\tilde{u}^2}\right] \beta_F.
\end{align}
Knowing the tensor structure, we can compute the coefficients by choosing spherical coordinates where the $z$ axis is aligned with the tilt, $\tilde{u}_i = (0,0,\tilde{u})$, so that $\beta=\beta'^{ab}\delta^{ab}$ and $\beta_F$ can be computed as $\beta_F = (\beta-3\beta'^{zz})/2$. 
Note that
\begin{equation}
\label{eq:fcomma}
f_{n,a} = \partial_{k_a} \Theta(\mu - \epsilon_k) = v^a \partial_\epsilon \Theta(\mu - \epsilon) = - v^a \delta(\mu - \epsilon),
\end{equation}
where $v^a =  \hat k^a +\tilde{u}^a$ and $\epsilon_k =  k + \tilde{u} k \cos \theta$. 

The computation of $\beta'^{ab}$ can be done as follows

\begin{align}
\beta'^{ab} &= \frac{i \pi C}{4} \int \frac{k^2 dk d\Omega}{(2\pi)^3} \Theta[(1+\tilde{u}\cos \theta)k-\mu] \frac{k^ak^b}{k^4}  \delta(2k-\omega) \\
& = {\frac{i \pi C}{8} \int \frac{d\theta \sin \theta}{(2\pi)^2} \Theta[(1+\tilde{u}\cos \theta)\omega/2-\mu] \hat k^a \hat k^b}.
\end{align}

To evaluate this integral define $a = (\tfrac{2\mu}{\omega}-1)/\tilde{u}$ and $g_1(\omega) = a\Theta(1+a)\Theta(1-a) + \Theta(a-1)-\Theta(-a-1)$. Changing variables to $x = \cos \theta$ we have

\begin{align}
\beta &={ -\frac{i \pi C}{8} \int_{g_1}^1 \frac{dx}{(2\pi)^2} = -\frac{i \pi C}{8}  \frac{(1-g_1(\omega))}{(2\pi)^2}}, \\
\beta'^{zz} &= { -\frac{i \pi C}{8} \int_{g_1}^1 \frac{dx}{(2\pi)^2}x^2 = -\frac{i \pi C}{8}\frac{(1-g^3_1(\omega))/3}{(2\pi)^2}},
\end{align}
so {$\beta_F = -\tfrac{i \pi C}{8}  \tfrac{(-g_1(\omega)+g^3_1(\omega))}{(2\pi)^2}$}. Note that when $\tilde{u}\to 0$ then $g_1(\omega)\to -1$ for $\omega>2\mu$ and we get $\beta = -\beta_0/4$ with $\beta_0=\tfrac{i\pi e^3}{h^2}$. Together with the factor 4 in Eq.~\eqref{eq:reftotalcurrentapptwoindex} the trace of the interband DFG is quantized to $\beta_0$.

Along similar lines the computation of $\gamma_1^{ab}$ reduces to
\begin{align}
{\gamma}'^{ab}_1= -\frac{iC}{4\omega} \int \frac{dk d\Omega}{(2\pi)^3}  ( \hat k^a +\tilde{u}^a) \hat k^b  \frac{\delta(k-k(\mu,\theta))}{|(1 +\tilde{u} \cos \theta)|} \frac{\omega}{4k^2 - \omega^2} \\
=  -\frac{iC}{4\omega} \int \frac{d\Omega}{(2\pi)^3} ( \hat k^a +\tilde{u}^a) \hat k^b  \frac{1}{|(1 + \tilde{u} \cos \theta)|} \frac{\omega}{\frac{4\mu^2}{(1 + \tilde{u} \cos \theta)^2} - \omega^2}.
\end{align}
The integrals can be evaluated in \texttt{Mathematica} resulting in the expressions provided in the main text:
\begin{align}
\gamma_1 &= -\dfrac{\beta_0}{2\pi} \frac{-\tilde{u} \omega +\mu\; g_2(\omega)}{{\tilde{u}} \omega^2},\\
\gamma'^{zz}_1 &= -\dfrac{\beta_0}{4\pi}\frac{1}{\tilde{u}^3 \omega^2} \Big[ 
2\omega\left(\tilde{u}\omega-\tilde{u}^3\omega+(\tilde{u}^2-2)\mu g_2(\omega)\right)
-(4\mu^2-(-1+\tilde{u}^2)\omega^2)g_3(\omega) \Big],
\end{align}
where $g_2(\omega) = {\rm arctanh}\left( \frac{4\tilde{u} \mu \omega}{4\mu^2+(\tilde{u}^2-1)\omega^2}\right)$ and $g_3(\omega) = {\rm arctanh}\left(\frac{2\tilde{u} \omega^2}{-4\mu^2+(\tilde{u}^2+1)\omega^2}\right)$. Using $\gamma_{1,F} = (\gamma_{1}-3\gamma'^{zz}_{1})/2$ we obtain
\begin{align}
\gamma_{1,F} &= -\dfrac{\beta_0}{8\pi}\frac{1}{\tilde{u}^3 \omega^2} \Big[ 
(-6\tilde{u}+4\tilde{u}^3)\omega^2
-(-12+4\tilde{u}^2)\mu\omega g_2(\omega)
+(12\mu^2+3\omega^2-3\tilde{u}^2\omega^2)g_3(\omega),
\Big]
\end{align}
which is given in the main text. It is also possible to check that these expressions indeed coincide with those obtained with the Kramers-Kronig transformations, given by Eq.~\eqref{KKtouse}.

Finally, the computation of the semiclassical contribution $\gamma_2^{ab}$ is as follows

\begin{equation}
f_{n,a} = -v^a\delta[\mu - k(1 + \tilde{u} \cos \theta)] = -\frac{v^a\delta(k-k(\mu,\theta))}{|(1 + \tilde{u} \cos \theta)|},
\end{equation} 
where $k(\mu,\theta) = \mu/(1 + \tilde{u} \cos \theta)$. Using $\Omega^a = \tfrac{k^a}{2k^3}$ we get
\begin{align}
\gamma'^{ab}_2&= \frac{iC}{4\omega} \int \frac{dk d\Omega}{(2\pi)^3} ( ( \hat k^a +\tilde{u}^a) \hat k^b - \delta^{ab} (1 + \tilde{u}^c\hat k^c)) \frac{\delta(k-k(\mu,\theta))}{|(1 + \tilde{u} \cos \theta)|} \\
&= \frac{iC}{4\omega} \int \frac{d\Omega}{(2\pi)^3} \frac{( \hat k^a +\tilde{u}^a) \hat k^b - \delta^{ab} (1 + \tilde{u}^c\hat k^c)}{|(1 + \tilde{u} \cos \theta)|},
\end{align}
where we have assumed $\tilde{u}<v_F$. Decomposing intro trace and traceless parts with Eq. \eqref{trdecomp}, we only need to compute
\begin{align}
{\gamma}_2 &= -\frac{iC}{2\omega} \int \frac{d\theta \sin \theta}{(2\pi)^2} \frac{1 +\tilde{u} \cos \theta}{|(1 + \tilde{u} \cos \theta)|} = -\frac{i C}{4\pi^2 \omega},\\
{\gamma}'^{zz}_2&= -\frac{iC}{4\omega} \int \frac{d\theta \sin \theta}{(2\pi)^2} \frac{(1-\cos^2\theta)}{|(1 + \tilde{u} \cos \theta)|} = -\frac{iC(\tilde{u}+(\tilde{u}^2-1){\rm arctanh}(\tilde{u}))}{8\pi^2\tilde{u}^3\omega}.
\end{align}
Using $\gamma_{2,F} = (\gamma_{2}-3\gamma'^{zz}_{2})/2$ we obtain
\begin{equation}
   \gamma_{2,F} = \frac{-iC}{16\pi^2 \omega \tilde{u}^3}\left[2\tilde{u}^3- 3(\tilde{u}+(\tilde{u}^2-1) {\rm arctanh}(\tilde{u})\right].
\end{equation}

\section{Weyl node pinned at a TRIM}

In this section we consider a Weyl node near the Fermi level and pinned to a time-reversal invariant momentum (TRIM) by time-reversal symmetry. 
These nodes appear in chiral structures~\cite{KramersWeyl} and their partners of opposite chirality are generally far away in energy and hence Pauli blocked. A single Weyl node can therefore account for the response of the whole crystal, if quadratic corrections to the dispersion are included.  
Due to time-reversal symmetry, $u_i =0$ and the only quadratic term allowed is proportional to the identity 
\begin{equation}
H_{\mathrm{TRIM}} = \sigma_i v_{ij} k_j + k_i A_{ij} k_j. \label{HTRIM}
\end{equation}
For simplicity we now restrict the example to chiral materials with point group $D_{3}$ such as TlTe$_2$O$_6$~\cite{KramersWeyl}, where the only non-zero parameters are $v_{xx} = v_{yy}\equiv v_{\parallel}$, $v_{zz} \equiv v_{\perp}$ and $A_{xx} = A_{yy} \equiv A_{\parallel}$, $A_{zz} \equiv v_\perp$.
This model is also applicable to the conduction band of elemental Te around the H point (not a TRIM) because the spin-dependent quadratic terms are negligible \cite{IP75}. 

Here we computate the traces $\beta$ and $\gamma$ for the Hamiltonian in Eq. \eqref{HTRIM}. The energies after the change of variables are given by $\epsilon_{k'} = \pm k' + a k'^2 + b k'^2_z$ with $a = A_\parallel/v_\parallel^2$ and $b = A_\perp/v_\perp^2-A_\parallel/v_\parallel^2$. We assume that $a>0$ and $b>0$. The only modification from the quadratic term occurs in the Fermi functions, which are $f_1 = \Theta(-k' + a k'^2 + b k'^2_z -\mu)$ and $f_2=\Theta(k' + a k'^2 + b k'^2_z -\mu)$. 

\begin{align}
\beta' &= \frac{i \pi C}{4} \int \frac{dk d\Omega}{(2\pi)^3} \left( \Theta(-k' + a k'^2 + b k'^2_z -\mu)-\Theta(k' + a k'^2 + b k'^2_z -\mu)\right)  \delta(2k-\omega) \\
& = \frac{i \pi C}{8} \int \frac{d\theta \sin \theta}{(2\pi)^2} \left( \Theta(-\omega/2 + (a  + b \cos^2 \theta)\omega^2/4 -\mu)-\Theta(\omega/2 + (a  + b \cos^2 \theta)\omega^2/4 -\mu)\right) .
\end{align}
Defining $c_{\pm} = (\pm2/\omega + a - 4\mu/\omega^2)/b$ and the functions $f_{\pm}= \Theta(c_\pm +1)\Theta(-c_\pm)c_\pm+\Theta(c_\pm)$ and changing variables to $x= \cos \theta$ we have 
\begin{align}
\beta' &= - \frac{i \pi C}{8(2\pi)^2} \int_{-1}^1 dx \left( \theta(c_- + x^2)-\theta(c_+ + x^2)\right) = - \frac{i \pi C}{8(2\pi)^2} \left( \int_{-|f_-|^{1/2}}^{|f_-|^{1/2}} dx - \int_{-|f_+|^{1/2}}^{|f_+|^{1/2}} dx \right) \\
 &= - \frac{i \pi C}{4(2\pi)^2} (|f_-|^{1/2}-|f_+|^{1/2}).
\end{align}

To obtain $\gamma_1$ we use the Kramers-Kronig formula \eqref{KKtouse}

\begin{align}
\gamma_1 =  - \frac{i \pi C}{2(2\pi)^2} \frac{2\omega}{\pi}\int_0^\infty d\omega' \frac{ \partial_{\omega'} (|f_-|^{1/2}-|f_+|^{1/2})}{\omega'^2 - \omega^2},
\end{align}
and change variables to $y_{\pm} = |f_\pm(\omega)|^{1/2}$, $dy_\pm = \partial_{\omega'} |f_\pm(\omega')|^{1/2} d\omega'$ to get 
\begin{align}
\gamma_1 =  - \frac{i C}{(2\pi)^2} \omega \left[\int_0^1 dy_- \frac{1}{\omega'^2(y_-) - \omega^2} - \int_0^1 dy_+\frac{1}{\omega'^2(y_+) - \omega^2}\right],
\end{align}
where $\omega'^2(y_\pm) = \left(\frac{1\pm\sqrt{1+4\mu(a+b y_\pm)}}{a+b y_\pm}\right)^2$. The integrals give 

\begin{align}
\gamma_1 =  - \frac{i C}{(2\pi)^2} \frac{2}{b\omega^2} \left( g_+ ({\rm arctanh} \tfrac{B}{g_+} - {\rm arctanh} \tfrac{A}{g_+}) + g_- ({\rm arctanh} \tfrac{A}{g_-} - {\rm arctanh} \tfrac{B}{g_-})\right),
\end{align}
where $A = \sqrt{1+4\mu a}$, $B = \sqrt{1+4\mu(a+b)}$, $g_\pm = \tfrac{4\mu}{\omega}\pm1$. 

We note that for $H_{\mathrm{TRIM}}$ $\gamma_2^{ab}$ was computed ab-initio for Te in Ref. \cite{SouzaTe} and 
the injection piece was computed in~\cite{Ma17}. 

\subsection{Threefold and fourfold fermion}

For a symmetric threefold fermion where the lowest bands (1,2) are completely filled,
the matrix elements restrict the transitions to be zero unless $m=3$ and $n=2$.
Evaluating the integrals as in the previous sections with $\Delta^{a}_{23}=-\partial_{k_{a}}\omega_3= -v_F k^{a}/k$, $\Omega^{b}_3=k^{b}/k^3$, $\Omega^{b}_2=0$ and Eq.~\eqref{eq:fcomma} we get

\begin{align}
\beta^{ab} &= 
-\dfrac{1}{3}\dfrac{\beta_0}{2}\delta_{ab}\Theta(\omega-\mu_{3f}), \\
\gamma_1^{ab} &= -\dfrac{1}{3\pi}\beta_0\frac{\omega}{\mu_{3f}^2-\omega^2}\delta_{ab}, \\
\gamma_2^{ab} 
&=-\frac{2\beta_0}{3\pi\omega} \delta^{ab},
\end{align}

where $\beta_0 = i C/(4\pi) = i\pi e^3/h$ is the universal circular photogalvanic quantum. 
A symmetric fourfold fermion is composed by two Weyls with the same chirality. The analytic results are given by

\begin{align}
\beta^{ab} 
&= -\dfrac{1}{3}\dfrac{\beta_0}{2}\delta_{ab}\Theta(\omega-2\mu_{4f}),  \\
\gamma_1^{ab} 
&= -\dfrac{1}{3\pi}\beta_0\delta_{ab}\frac{\omega}{(2\mu_{4f})^2-\omega^2}, \\
\gamma_2^{ab} 
&= -\dfrac{2\beta_0}{3\pi\omega}\delta_{ab},
\end{align}
Fig.~\ref{fig2} in the main text is computed by summing the contributions of a threefold with chemical potential $\mu_{3f}$ and a fourfold with chemical potential $\mu_{4f}$ of opposite chiralities.

\section{Details of the ab-initio calculation}

TaAs has space group \cite{TaAsLv} $I4_1 md$ ($\#$109), with point group $4mm$ ($C_{4v}$). The two index circular photogalvanic tensors have the same symmetry as the gyrotropic tensor, which in $C_{4v}$ only has components $\beta^{xy} = - \beta^{yx}$, and similar for $\gamma^{ab}$ \cite{SturmanBook}. To calculate the free-carrier current and injection current in TaAs, we obtain the density-functional theory (DFT) Bloch wave functions from the Full-Potential Local-Orbital program (FPLO) \cite{koepernik1999full} within the generalized gradient approximation (GGA) \cite{perdew1996}. By projecting the Bloch wave functions onto Wannier functions, we obtain a tight-binding Hamiltonian with 32 bands, which we use for efficient evaluation of the photocurrent. 
To implement the circular DFG integrals the Brillouin zone was sampled by $k$-grids from $200\times200\times200$ to $960\times960\times960$ \cite{ZSY18}. Satisfactory convergence was achieved for a k-grid of size $400\times400\times400$. 
To compute $\gamma^{ab}_1$ from ab-initio, it is better to go back to the Kramers-Kronig formula because the Fermi surface formula requires the derivative of Fermi-Dirac distribution function on momentum space, which is numerically very unstable and hard to converge.  $\gamma^{ab}_1$ is obtained via Eq. \eqref{KKtouse} with the use of Kramers-Kronig transformations.

\end{widetext}

\begin{thebibliography}{70}%
\makeatletter
\providecommand \@ifxundefined [1]{%
 \@ifx{#1\undefined}
}%
\providecommand \@ifnum [1]{%
 \ifnum #1\expandafter \@firstoftwo
 \else \expandafter \@secondoftwo
 \fi
}%
\providecommand \@ifx [1]{%
 \ifx #1\expandafter \@firstoftwo
 \else \expandafter \@secondoftwo
 \fi
}%
\providecommand \natexlab [1]{#1}%
\providecommand \enquote  [1]{``#1''}%
\providecommand \bibnamefont  [1]{#1}%
\providecommand \bibfnamefont [1]{#1}%
\providecommand \citenamefont [1]{#1}%
\providecommand \href@noop [0]{\@secondoftwo}%
\providecommand \href [0]{\begingroup \@sanitize@url \@href}%
\providecommand \@href[1]{\@@startlink{#1}\@@href}%
\providecommand \@@href[1]{\endgroup#1\@@endlink}%
\providecommand \@sanitize@url [0]{\catcode `\\12\catcode `\$12\catcode
  `\&12\catcode `\#12\catcode `\^12\catcode `\_12\catcode `\%12\relax}%
\providecommand \@@startlink[1]{}%
\providecommand \@@endlink[0]{}%
\providecommand \url  [0]{\begingroup\@sanitize@url \@url }%
\providecommand \@url [1]{\endgroup\@href {#1}{\urlprefix }}%
\providecommand \urlprefix  [0]{URL }%
\providecommand \Eprint [0]{\href }%
\providecommand \doibase [0]{http://dx.doi.org/}%
\providecommand \selectlanguage [0]{\@gobble}%
\providecommand \bibinfo  [0]{\@secondoftwo}%
\providecommand \bibfield  [0]{\@secondoftwo}%
\providecommand \translation [1]{[#1]}%
\providecommand \BibitemOpen [0]{}%
\providecommand \bibitemStop [0]{}%
\providecommand \bibitemNoStop [0]{.\EOS\space}%
\providecommand \EOS [0]{\spacefactor3000\relax}%
\providecommand \BibitemShut  [1]{\csname bibitem#1\endcsname}%
\let\auto@bib@innerbib\@empty
\bibitem [{\citenamefont {de~Juan}\ \emph {et~al.}(2017)\citenamefont
  {de~Juan}, \citenamefont {Grushin}, \citenamefont {Morimoto},\ and\
  \citenamefont {Moore}}]{deJuan17}%
  \BibitemOpen
  \bibfield  {author} {\bibinfo {author} {\bibfnamefont {F.}~\bibnamefont
  {de~Juan}}, \bibinfo {author} {\bibfnamefont {A.}~\bibnamefont {Grushin}},
  \bibinfo {author} {\bibfnamefont {T.}~\bibnamefont {Morimoto}}, \ and\
  \bibinfo {author} {\bibfnamefont {J.}~\bibnamefont {Moore}},\ }\href
  {\doibase doi:10.1038/ncomms15995} {\bibfield  {journal} {\bibinfo  {journal}
  {Nat. Commun.}\ }\textbf {\bibinfo {volume} {8}},\ \bibinfo {pages} {15995}
  (\bibinfo {year} {2017})}\BibitemShut {NoStop}%
\bibitem [{\citenamefont {Flicker}\ \emph {et~al.}(2018)\citenamefont
  {Flicker}, \citenamefont {de~Juan}, \citenamefont {Bradlyn}, \citenamefont
  {Morimoto}, \citenamefont {Vergniory},\ and\ \citenamefont
  {Grushin}}]{Flicker2018}%
  \BibitemOpen
  \bibfield  {author} {\bibinfo {author} {\bibfnamefont {F.}~\bibnamefont
  {Flicker}}, \bibinfo {author} {\bibfnamefont {F.}~\bibnamefont {de~Juan}},
  \bibinfo {author} {\bibfnamefont {B.}~\bibnamefont {Bradlyn}}, \bibinfo
  {author} {\bibfnamefont {T.}~\bibnamefont {Morimoto}}, \bibinfo {author}
  {\bibfnamefont {M.~G.}\ \bibnamefont {Vergniory}}, \ and\ \bibinfo {author}
  {\bibfnamefont {A.~G.}\ \bibnamefont {Grushin}},\ }\href {\doibase
  10.1103/PhysRevB.98.155145} {\bibfield  {journal} {\bibinfo  {journal} {Phys.
  Rev. B}\ }\textbf {\bibinfo {volume} {98}},\ \bibinfo {pages} {155145}
  (\bibinfo {year} {2018})}\BibitemShut {NoStop}%
\bibitem [{\citenamefont {Hirayama}\ \emph {et~al.}(2015)\citenamefont
  {Hirayama}, \citenamefont {Okugawa}, \citenamefont {Ishibashi}, \citenamefont
  {Murakami},\ and\ \citenamefont {Miyake}}]{HOI15}%
  \BibitemOpen
  \bibfield  {author} {\bibinfo {author} {\bibfnamefont {M.}~\bibnamefont
  {Hirayama}}, \bibinfo {author} {\bibfnamefont {R.}~\bibnamefont {Okugawa}},
  \bibinfo {author} {\bibfnamefont {S.}~\bibnamefont {Ishibashi}}, \bibinfo
  {author} {\bibfnamefont {S.}~\bibnamefont {Murakami}}, \ and\ \bibinfo
  {author} {\bibfnamefont {T.}~\bibnamefont {Miyake}},\ }\href {\doibase
  10.1103/PhysRevLett.114.206401} {\bibfield  {journal} {\bibinfo  {journal}
  {Phys. Rev. Lett.}\ }\textbf {\bibinfo {volume} {114}},\ \bibinfo {pages}
  {206401} (\bibinfo {year} {2015})}\BibitemShut {NoStop}%
\bibitem [{\citenamefont {Huang}\ \emph {et~al.}(2016)\citenamefont {Huang},
  \citenamefont {Xu}, \citenamefont {Belopolski}, \citenamefont {Lee},
  \citenamefont {Chang}, \citenamefont {Chang}, \citenamefont {Wang},
  \citenamefont {Alidoust}, \citenamefont {Bian}, \citenamefont {Neupane},
  \citenamefont {Sanchez}, \citenamefont {Zheng}, \citenamefont {Jeng},
  \citenamefont {Bansil}, \citenamefont {Neupert}, \citenamefont {Lin},\ and\
  \citenamefont {Hasan}}]{HXB16}%
  \BibitemOpen
  \bibfield  {author} {\bibinfo {author} {\bibfnamefont {S.-M.}\ \bibnamefont
  {Huang}}, \bibinfo {author} {\bibfnamefont {S.-Y.}\ \bibnamefont {Xu}},
  \bibinfo {author} {\bibfnamefont {I.}~\bibnamefont {Belopolski}}, \bibinfo
  {author} {\bibfnamefont {C.-C.}\ \bibnamefont {Lee}}, \bibinfo {author}
  {\bibfnamefont {G.}~\bibnamefont {Chang}}, \bibinfo {author} {\bibfnamefont
  {T.-R.}\ \bibnamefont {Chang}}, \bibinfo {author} {\bibfnamefont
  {B.}~\bibnamefont {Wang}}, \bibinfo {author} {\bibfnamefont {N.}~\bibnamefont
  {Alidoust}}, \bibinfo {author} {\bibfnamefont {G.}~\bibnamefont {Bian}},
  \bibinfo {author} {\bibfnamefont {M.}~\bibnamefont {Neupane}}, \bibinfo
  {author} {\bibfnamefont {D.}~\bibnamefont {Sanchez}}, \bibinfo {author}
  {\bibfnamefont {H.}~\bibnamefont {Zheng}}, \bibinfo {author} {\bibfnamefont
  {H.-T.}\ \bibnamefont {Jeng}}, \bibinfo {author} {\bibfnamefont
  {A.}~\bibnamefont {Bansil}}, \bibinfo {author} {\bibfnamefont
  {T.}~\bibnamefont {Neupert}}, \bibinfo {author} {\bibfnamefont
  {H.}~\bibnamefont {Lin}}, \ and\ \bibinfo {author} {\bibfnamefont {M.~Z.}\
  \bibnamefont {Hasan}},\ }\href@noop {} {\bibfield  {journal} {\bibinfo
  {journal} {Proc. Nat. Acad. Sci.}\ }\textbf {\bibinfo {volume} {113}},\
  \bibinfo {pages} {1180} (\bibinfo {year} {2016})}\BibitemShut {NoStop}%
\bibitem [{\citenamefont {Chang}\ \emph {et~al.}(2018)\citenamefont {Chang},
  \citenamefont {Wieder}, \citenamefont {Schindler}, \citenamefont {Sanchez},
  \citenamefont {Belopolski}, \citenamefont {Huang}, \citenamefont {Singh},
  \citenamefont {Wu}, \citenamefont {Chang}, \citenamefont {Neupert},
  \citenamefont {Xu}, \citenamefont {Lin},\ and\ \citenamefont
  {Hasan}}]{KramersWeyl}%
  \BibitemOpen
  \bibfield  {author} {\bibinfo {author} {\bibfnamefont {G.}~\bibnamefont
  {Chang}}, \bibinfo {author} {\bibfnamefont {B.~J.}\ \bibnamefont {Wieder}},
  \bibinfo {author} {\bibfnamefont {F.}~\bibnamefont {Schindler}}, \bibinfo
  {author} {\bibfnamefont {D.~S.}\ \bibnamefont {Sanchez}}, \bibinfo {author}
  {\bibfnamefont {I.}~\bibnamefont {Belopolski}}, \bibinfo {author}
  {\bibfnamefont {S.-M.}\ \bibnamefont {Huang}}, \bibinfo {author}
  {\bibfnamefont {B.}~\bibnamefont {Singh}}, \bibinfo {author} {\bibfnamefont
  {D.}~\bibnamefont {Wu}}, \bibinfo {author} {\bibfnamefont {T.-R.}\
  \bibnamefont {Chang}}, \bibinfo {author} {\bibfnamefont {T.}~\bibnamefont
  {Neupert}}, \bibinfo {author} {\bibfnamefont {S.-Y.}\ \bibnamefont {Xu}},
  \bibinfo {author} {\bibfnamefont {H.}~\bibnamefont {Lin}}, \ and\ \bibinfo
  {author} {\bibfnamefont {M.~Z.}\ \bibnamefont {Hasan}},\ }\href {\doibase
  10.1038/s41563-018-0169-3} {\bibfield  {journal} {\bibinfo  {journal} {Nat.
  Mater.}\ }\textbf {\bibinfo {volume} {17}},\ \bibinfo {pages} {978} (\bibinfo
  {year} {2018})}\BibitemShut {NoStop}%
\bibitem [{\citenamefont {Laman}\ \emph {et~al.}(2005)\citenamefont {Laman},
  \citenamefont {Bieler},\ and\ \citenamefont {Van~Driel}}]{Laman05}%
  \BibitemOpen
  \bibfield  {author} {\bibinfo {author} {\bibfnamefont {N.}~\bibnamefont
  {Laman}}, \bibinfo {author} {\bibfnamefont {M.}~\bibnamefont {Bieler}}, \
  and\ \bibinfo {author} {\bibfnamefont {H.}~\bibnamefont {Van~Driel}},\
  }\href@noop {} {\bibfield  {journal} {\bibinfo  {journal} {Journal of applied
  physics}\ }\textbf {\bibinfo {volume} {98}},\ \bibinfo {pages} {103507}
  (\bibinfo {year} {2005})}\BibitemShut {NoStop}%
\bibitem [{\citenamefont {Bieler}\ \emph {et~al.}(2006)\citenamefont {Bieler},
  \citenamefont {Pierz},\ and\ \citenamefont {Siegner}}]{Bieler06}%
  \BibitemOpen
  \bibfield  {author} {\bibinfo {author} {\bibfnamefont {M.}~\bibnamefont
  {Bieler}}, \bibinfo {author} {\bibfnamefont {K.}~\bibnamefont {Pierz}}, \
  and\ \bibinfo {author} {\bibfnamefont {U.}~\bibnamefont {Siegner}},\
  }\href@noop {} {\bibfield  {journal} {\bibinfo  {journal} {J. App. Phys.}\
  }\textbf {\bibinfo {volume} {100}},\ \bibinfo {pages} {083710} (\bibinfo
  {year} {2006})}\BibitemShut {NoStop}%
\bibitem [{\citenamefont {Bas}\ \emph {et~al.}(2016)\citenamefont {Bas},
  \citenamefont {Muniz}, \citenamefont {Babakiray}, \citenamefont {Lederman},
  \citenamefont {Sipe},\ and\ \citenamefont {Bristow}}]{BMB16}%
  \BibitemOpen
  \bibfield  {author} {\bibinfo {author} {\bibfnamefont {D.~A.}\ \bibnamefont
  {Bas}}, \bibinfo {author} {\bibfnamefont {R.~A.}\ \bibnamefont {Muniz}},
  \bibinfo {author} {\bibfnamefont {S.}~\bibnamefont {Babakiray}}, \bibinfo
  {author} {\bibfnamefont {D.}~\bibnamefont {Lederman}}, \bibinfo {author}
  {\bibfnamefont {J.}~\bibnamefont {Sipe}}, \ and\ \bibinfo {author}
  {\bibfnamefont {A.~D.}\ \bibnamefont {Bristow}},\ }\href@noop {} {\bibfield
  {journal} {\bibinfo  {journal} {Opt. Express}\ }\textbf {\bibinfo {volume}
  {24}},\ \bibinfo {pages} {23583} (\bibinfo {year} {2016})}\BibitemShut
  {NoStop}%
\bibitem [{\citenamefont {Rees}\ \emph {et~al.}(2019)\citenamefont {Rees},
  \citenamefont {Manna}, \citenamefont {Lu}, \citenamefont {Morimoto},
  \citenamefont {Borrmann}, \citenamefont {Felser}, \citenamefont {Moore},
  \citenamefont {Torchinsky},\ and\ \citenamefont {Orenstein}}]{Rees19}%
  \BibitemOpen
  \bibfield  {author} {\bibinfo {author} {\bibfnamefont {D.}~\bibnamefont
  {Rees}}, \bibinfo {author} {\bibfnamefont {K.}~\bibnamefont {Manna}},
  \bibinfo {author} {\bibfnamefont {B.}~\bibnamefont {Lu}}, \bibinfo {author}
  {\bibfnamefont {T.}~\bibnamefont {Morimoto}}, \bibinfo {author}
  {\bibfnamefont {H.}~\bibnamefont {Borrmann}}, \bibinfo {author}
  {\bibfnamefont {C.}~\bibnamefont {Felser}}, \bibinfo {author} {\bibfnamefont
  {J.}~\bibnamefont {Moore}}, \bibinfo {author} {\bibfnamefont {D.~H.}\
  \bibnamefont {Torchinsky}}, \ and\ \bibinfo {author} {\bibfnamefont
  {J.}~\bibnamefont {Orenstein}},\ }\href@noop {} {\bibfield  {journal}
  {\bibinfo  {journal} {arXiv:1902.03230}\ } (\bibinfo {year}
  {2019})}\BibitemShut {NoStop}%
\bibitem [{\citenamefont {Sipe}\ and\ \citenamefont
  {Ghahramani}(1993)}]{GhahramaniSipe}%
  \BibitemOpen
  \bibfield  {author} {\bibinfo {author} {\bibfnamefont {J.~E.}\ \bibnamefont
  {Sipe}}\ and\ \bibinfo {author} {\bibfnamefont {E.}~\bibnamefont
  {Ghahramani}},\ }\href {\doibase 10.1103/PhysRevB.48.11705} {\bibfield
  {journal} {\bibinfo  {journal} {Phys. Rev. B}\ }\textbf {\bibinfo {volume}
  {48}},\ \bibinfo {pages} {11705} (\bibinfo {year} {1993})}\BibitemShut
  {NoStop}%
\bibitem [{\citenamefont {Aversa}\ and\ \citenamefont
  {Sipe}(1995)}]{AversaSipe}%
  \BibitemOpen
  \bibfield  {author} {\bibinfo {author} {\bibfnamefont {C.}~\bibnamefont
  {Aversa}}\ and\ \bibinfo {author} {\bibfnamefont {J.~E.}\ \bibnamefont
  {Sipe}},\ }\href {\doibase 10.1103/PhysRevB.52.14636} {\bibfield  {journal}
  {\bibinfo  {journal} {Phys. Rev. B}\ }\textbf {\bibinfo {volume} {52}},\
  \bibinfo {pages} {14636} (\bibinfo {year} {1995})}\BibitemShut {NoStop}%
\bibitem [{\citenamefont {Sipe}\ and\ \citenamefont
  {Shkrebtii}(2000)}]{SipeShkrebtii}%
  \BibitemOpen
  \bibfield  {author} {\bibinfo {author} {\bibfnamefont {J.~E.}\ \bibnamefont
  {Sipe}}\ and\ \bibinfo {author} {\bibfnamefont {A.~I.}\ \bibnamefont
  {Shkrebtii}},\ }\href {\doibase 10.1103/PhysRevB.61.5337} {\bibfield
  {journal} {\bibinfo  {journal} {Phys. Rev. B}\ }\textbf {\bibinfo {volume}
  {61}},\ \bibinfo {pages} {5337} (\bibinfo {year} {2000})}\BibitemShut
  {NoStop}%
\bibitem [{\citenamefont {Nastos}\ and\ \citenamefont
  {Sipe}(2006)}]{NastosSipe06}%
  \BibitemOpen
  \bibfield  {author} {\bibinfo {author} {\bibfnamefont {F.}~\bibnamefont
  {Nastos}}\ and\ \bibinfo {author} {\bibfnamefont {J.~E.}\ \bibnamefont
  {Sipe}},\ }\href {\doibase 10.1103/PhysRevB.74.035201} {\bibfield  {journal}
  {\bibinfo  {journal} {Phys. Rev. B}\ }\textbf {\bibinfo {volume} {74}},\
  \bibinfo {pages} {035201} (\bibinfo {year} {2006})}\BibitemShut {NoStop}%
\bibitem [{\citenamefont {Nastos}\ and\ \citenamefont
  {Sipe}(2010)}]{NastosSipe10}%
  \BibitemOpen
  \bibfield  {author} {\bibinfo {author} {\bibfnamefont {F.}~\bibnamefont
  {Nastos}}\ and\ \bibinfo {author} {\bibfnamefont {J.~E.}\ \bibnamefont
  {Sipe}},\ }\href {\doibase 10.1103/PhysRevB.82.235204} {\bibfield  {journal}
  {\bibinfo  {journal} {Phys. Rev. B}\ }\textbf {\bibinfo {volume} {82}},\
  \bibinfo {pages} {235204} (\bibinfo {year} {2010})}\BibitemShut {NoStop}%
\bibitem [{\citenamefont {Genkin}\ and\ \citenamefont
  {Mednis}(1968)}]{Genkin68}%
  \BibitemOpen
  \bibfield  {author} {\bibinfo {author} {\bibfnamefont {V.}~\bibnamefont
  {Genkin}}\ and\ \bibinfo {author} {\bibfnamefont {P.}~\bibnamefont
  {Mednis}},\ }\href@noop {} {\bibfield  {journal} {\bibinfo  {journal} {Sov.
  Phys. JETP}\ }\textbf {\bibinfo {volume} {27}},\ \bibinfo {pages} {609}
  (\bibinfo {year} {1968})}\BibitemShut {NoStop}%
\bibitem [{\citenamefont {Sodemann}\ and\ \citenamefont
  {Fu}(2015)}]{sodemannfu}%
  \BibitemOpen
  \bibfield  {author} {\bibinfo {author} {\bibfnamefont {I.}~\bibnamefont
  {Sodemann}}\ and\ \bibinfo {author} {\bibfnamefont {L.}~\bibnamefont {Fu}},\
  }\href {\doibase 10.1103/PhysRevLett.115.216806} {\bibfield  {journal}
  {\bibinfo  {journal} {Phys. Rev. Lett.}\ }\textbf {\bibinfo {volume} {115}},\
  \bibinfo {pages} {216806} (\bibinfo {year} {2015})}\BibitemShut {NoStop}%
\bibitem [{\citenamefont {Belinicher}\ \emph {et~al.}(1986)\citenamefont
  {Belinicher}, \citenamefont {Ivchenko},\ and\ \citenamefont
  {Pikus}}]{BelinicherTransient}%
  \BibitemOpen
  \bibfield  {author} {\bibinfo {author} {\bibfnamefont {V.}~\bibnamefont
  {Belinicher}}, \bibinfo {author} {\bibfnamefont {E.}~\bibnamefont
  {Ivchenko}}, \ and\ \bibinfo {author} {\bibfnamefont {G.}~\bibnamefont
  {Pikus}},\ }\href@noop {} {\bibfield  {journal} {\bibinfo  {journal} {Sov.
  Phys. Semicond.}\ }\textbf {\bibinfo {volume} {20}},\ \bibinfo {pages} {558}
  (\bibinfo {year} {1986})}\BibitemShut {NoStop}%
\bibitem [{\citenamefont {Belinicher}\ \emph {et~al.}(1982)\citenamefont
  {Belinicher}, \citenamefont {Ivchenko},\ and\ \citenamefont
  {Sturman}}]{BelinicherKinetic}%
  \BibitemOpen
  \bibfield  {author} {\bibinfo {author} {\bibfnamefont {V.}~\bibnamefont
  {Belinicher}}, \bibinfo {author} {\bibfnamefont {E.}~\bibnamefont
  {Ivchenko}}, \ and\ \bibinfo {author} {\bibfnamefont {B.}~\bibnamefont
  {Sturman}},\ }\href@noop {} {\bibfield  {journal} {\bibinfo  {journal} {Zh.
  Eksp. Teor. Fiz.}\ }\textbf {\bibinfo {volume} {83}},\ \bibinfo {pages} {649}
  (\bibinfo {year} {1982})}\BibitemShut {NoStop}%
\bibitem [{\citenamefont {Yang}\ \emph {et~al.}(2017)\citenamefont {Yang},
  \citenamefont {Burch},\ and\ \citenamefont {Ran}}]{Ran17}%
  \BibitemOpen
  \bibfield  {author} {\bibinfo {author} {\bibfnamefont {X.}~\bibnamefont
  {Yang}}, \bibinfo {author} {\bibfnamefont {K.}~\bibnamefont {Burch}}, \ and\
  \bibinfo {author} {\bibfnamefont {Y.}~\bibnamefont {Ran}},\ }\href@noop {}
  {\bibfield  {journal} {\bibinfo  {journal} {arXiv preprint arXiv:1712.09363}\
  } (\bibinfo {year} {2017})}\BibitemShut {NoStop}%
\bibitem [{\citenamefont {Zhang}\ \emph
  {et~al.}(2018{\natexlab{a}})\citenamefont {Zhang}, \citenamefont {Ishizuka},
  \citenamefont {van~den Brink}, \citenamefont {Felser}, \citenamefont {Yan},\
  and\ \citenamefont {Nagaosa}}]{Zhang18}%
  \BibitemOpen
  \bibfield  {author} {\bibinfo {author} {\bibfnamefont {Y.}~\bibnamefont
  {Zhang}}, \bibinfo {author} {\bibfnamefont {H.}~\bibnamefont {Ishizuka}},
  \bibinfo {author} {\bibfnamefont {J.}~\bibnamefont {van~den Brink}}, \bibinfo
  {author} {\bibfnamefont {C.}~\bibnamefont {Felser}}, \bibinfo {author}
  {\bibfnamefont {B.}~\bibnamefont {Yan}}, \ and\ \bibinfo {author}
  {\bibfnamefont {N.}~\bibnamefont {Nagaosa}},\ }\href@noop {} {\bibfield
  {journal} {\bibinfo  {journal} {Physical Review B}\ }\textbf {\bibinfo
  {volume} {97}},\ \bibinfo {pages} {241118} (\bibinfo {year}
  {2018}{\natexlab{a}})}\BibitemShut {NoStop}%
\bibitem [{\citenamefont {Ma}\ \emph {et~al.}(2017)\citenamefont {Ma},
  \citenamefont {Xu}, \citenamefont {Chan}, \citenamefont {Zhang},
  \citenamefont {Chang}, \citenamefont {Lin}, \citenamefont {Xie},
  \citenamefont {Palacios}, \citenamefont {Lin}, \citenamefont {Jia} \emph
  {et~al.}}]{Ma17}%
  \BibitemOpen
  \bibfield  {author} {\bibinfo {author} {\bibfnamefont {Q.}~\bibnamefont
  {Ma}}, \bibinfo {author} {\bibfnamefont {S.-Y.}\ \bibnamefont {Xu}}, \bibinfo
  {author} {\bibfnamefont {C.-K.}\ \bibnamefont {Chan}}, \bibinfo {author}
  {\bibfnamefont {C.-L.}\ \bibnamefont {Zhang}}, \bibinfo {author}
  {\bibfnamefont {G.}~\bibnamefont {Chang}}, \bibinfo {author} {\bibfnamefont
  {Y.}~\bibnamefont {Lin}}, \bibinfo {author} {\bibfnamefont {W.}~\bibnamefont
  {Xie}}, \bibinfo {author} {\bibfnamefont {T.}~\bibnamefont {Palacios}},
  \bibinfo {author} {\bibfnamefont {H.}~\bibnamefont {Lin}}, \bibinfo {author}
  {\bibfnamefont {S.}~\bibnamefont {Jia}},  \emph {et~al.},\ }\href@noop {}
  {\bibfield  {journal} {\bibinfo  {journal} {Nature Physics}\ }\textbf
  {\bibinfo {volume} {13}},\ \bibinfo {pages} {842} (\bibinfo {year}
  {2017})}\BibitemShut {NoStop}%
\bibitem [{\citenamefont {Osterhoudt}\ \emph {et~al.}(2019)\citenamefont
  {Osterhoudt}, \citenamefont {Diebel}, \citenamefont {Gray}, \citenamefont
  {Yang}, \citenamefont {Stanco}, \citenamefont {Huang}, \citenamefont {Shen},
  \citenamefont {Ni}, \citenamefont {Moll}, \citenamefont {Ran},\ and\
  \citenamefont {Burch}}]{Burch17}%
  \BibitemOpen
  \bibfield  {author} {\bibinfo {author} {\bibfnamefont {G.~B.}\ \bibnamefont
  {Osterhoudt}}, \bibinfo {author} {\bibfnamefont {L.~K.}\ \bibnamefont
  {Diebel}}, \bibinfo {author} {\bibfnamefont {M.~J.}\ \bibnamefont {Gray}},
  \bibinfo {author} {\bibfnamefont {X.}~\bibnamefont {Yang}}, \bibinfo {author}
  {\bibfnamefont {J.}~\bibnamefont {Stanco}}, \bibinfo {author} {\bibfnamefont
  {X.}~\bibnamefont {Huang}}, \bibinfo {author} {\bibfnamefont
  {B.}~\bibnamefont {Shen}}, \bibinfo {author} {\bibfnamefont {N.}~\bibnamefont
  {Ni}}, \bibinfo {author} {\bibfnamefont {P.~J.~W.}\ \bibnamefont {Moll}},
  \bibinfo {author} {\bibfnamefont {Y.}~\bibnamefont {Ran}}, \ and\ \bibinfo
  {author} {\bibfnamefont {K.~S.}\ \bibnamefont {Burch}},\ }\href
  {https://doi.org/10.1038/s41563-019-0297-4} {\bibfield  {journal} {\bibinfo
  {journal} {Nature materials}\ }\textbf {\bibinfo {volume} {18}},\ \bibinfo
  {pages} {471} (\bibinfo {year} {2019})}\BibitemShut {NoStop}%
\bibitem [{\citenamefont {Wu}\ \emph {et~al.}(2017)\citenamefont {Wu},
  \citenamefont {Patankar}, \citenamefont {Morimoto}, \citenamefont {Nair},
  \citenamefont {Thewalt}, \citenamefont {Little}, \citenamefont {Analytis},
  \citenamefont {Moore},\ and\ \citenamefont {Orenstein}}]{Wu17}%
  \BibitemOpen
  \bibfield  {author} {\bibinfo {author} {\bibfnamefont {L.}~\bibnamefont
  {Wu}}, \bibinfo {author} {\bibfnamefont {S.}~\bibnamefont {Patankar}},
  \bibinfo {author} {\bibfnamefont {T.}~\bibnamefont {Morimoto}}, \bibinfo
  {author} {\bibfnamefont {N.~L.}\ \bibnamefont {Nair}}, \bibinfo {author}
  {\bibfnamefont {E.}~\bibnamefont {Thewalt}}, \bibinfo {author} {\bibfnamefont
  {A.}~\bibnamefont {Little}}, \bibinfo {author} {\bibfnamefont {J.~G.}\
  \bibnamefont {Analytis}}, \bibinfo {author} {\bibfnamefont {J.~E.}\
  \bibnamefont {Moore}}, \ and\ \bibinfo {author} {\bibfnamefont
  {J.}~\bibnamefont {Orenstein}},\ }\href@noop {} {\bibfield  {journal}
  {\bibinfo  {journal} {Nature Physics}\ }\textbf {\bibinfo {volume} {13}},\
  \bibinfo {pages} {350} (\bibinfo {year} {2017})}\BibitemShut {NoStop}%
\bibitem [{\citenamefont {Patankar}\ \emph {et~al.}(2018)\citenamefont
  {Patankar}, \citenamefont {Wu}, \citenamefont {Lu}, \citenamefont {Rai},
  \citenamefont {Tran}, \citenamefont {Morimoto}, \citenamefont {Parker},
  \citenamefont {Grushin}, \citenamefont {Nair}, \citenamefont {Analytis},
  \citenamefont {Moore}, \citenamefont {Orenstein},\ and\ \citenamefont
  {Torchinsky}}]{Patankar2018}%
  \BibitemOpen
  \bibfield  {author} {\bibinfo {author} {\bibfnamefont {S.}~\bibnamefont
  {Patankar}}, \bibinfo {author} {\bibfnamefont {L.}~\bibnamefont {Wu}},
  \bibinfo {author} {\bibfnamefont {B.}~\bibnamefont {Lu}}, \bibinfo {author}
  {\bibfnamefont {M.}~\bibnamefont {Rai}}, \bibinfo {author} {\bibfnamefont
  {J.~D.}\ \bibnamefont {Tran}}, \bibinfo {author} {\bibfnamefont
  {T.}~\bibnamefont {Morimoto}}, \bibinfo {author} {\bibfnamefont {D.~E.}\
  \bibnamefont {Parker}}, \bibinfo {author} {\bibfnamefont {A.~G.}\
  \bibnamefont {Grushin}}, \bibinfo {author} {\bibfnamefont {N.~L.}\
  \bibnamefont {Nair}}, \bibinfo {author} {\bibfnamefont {J.~G.}\ \bibnamefont
  {Analytis}}, \bibinfo {author} {\bibfnamefont {J.~E.}\ \bibnamefont {Moore}},
  \bibinfo {author} {\bibfnamefont {J.}~\bibnamefont {Orenstein}}, \ and\
  \bibinfo {author} {\bibfnamefont {D.~H.}\ \bibnamefont {Torchinsky}},\ }\href
  {\doibase 10.1103/PhysRevB.98.165113} {\bibfield  {journal} {\bibinfo
  {journal} {Phys. Rev. B}\ }\textbf {\bibinfo {volume} {98}},\ \bibinfo
  {pages} {165113} (\bibinfo {year} {2018})}\BibitemShut {NoStop}%
\bibitem [{\citenamefont {Ji}\ \emph {et~al.}(2018)\citenamefont {Ji},
  \citenamefont {Liu}, \citenamefont {Addison}, \citenamefont {Liu},
  \citenamefont {Yu}, \citenamefont {Gao}, \citenamefont {Liu}, \citenamefont
  {Rappe}, \citenamefont {Kane}, \citenamefont {Mele} \emph
  {et~al.}}]{Rappe18}%
  \BibitemOpen
  \bibfield  {author} {\bibinfo {author} {\bibfnamefont {Z.}~\bibnamefont
  {Ji}}, \bibinfo {author} {\bibfnamefont {G.}~\bibnamefont {Liu}}, \bibinfo
  {author} {\bibfnamefont {Z.}~\bibnamefont {Addison}}, \bibinfo {author}
  {\bibfnamefont {W.}~\bibnamefont {Liu}}, \bibinfo {author} {\bibfnamefont
  {P.}~\bibnamefont {Yu}}, \bibinfo {author} {\bibfnamefont {H.}~\bibnamefont
  {Gao}}, \bibinfo {author} {\bibfnamefont {Z.}~\bibnamefont {Liu}}, \bibinfo
  {author} {\bibfnamefont {A.~M.}\ \bibnamefont {Rappe}}, \bibinfo {author}
  {\bibfnamefont {C.~L.}\ \bibnamefont {Kane}}, \bibinfo {author}
  {\bibfnamefont {E.~J.}\ \bibnamefont {Mele}},  \emph {et~al.},\ }\href@noop
  {} {\bibfield  {journal} {\bibinfo  {journal} {arXiv:1802.04387}\ } (\bibinfo
  {year} {2018})}\BibitemShut {NoStop}%
\bibitem [{\citenamefont {Hosur}(2011)}]{Hosur11}%
  \BibitemOpen
  \bibfield  {author} {\bibinfo {author} {\bibfnamefont {P.}~\bibnamefont
  {Hosur}},\ }\href {\doibase 10.1103/PhysRevB.83.035309} {\bibfield  {journal}
  {\bibinfo  {journal} {Phys. Rev. B}\ }\textbf {\bibinfo {volume} {83}},\
  \bibinfo {pages} {035309} (\bibinfo {year} {2011})}\BibitemShut {NoStop}%
\bibitem [{\citenamefont {Morimoto}\ and\ \citenamefont
  {Nagaosa}(2016)}]{MN16}%
  \BibitemOpen
  \bibfield  {author} {\bibinfo {author} {\bibfnamefont {T.}~\bibnamefont
  {Morimoto}}\ and\ \bibinfo {author} {\bibfnamefont {N.}~\bibnamefont
  {Nagaosa}},\ }\href {\doibase 10.1126/sciadv.1501524} {\bibfield  {journal}
  {\bibinfo  {journal} {Science Advances}\ }\textbf {\bibinfo {volume} {2}},\
  \bibinfo {pages} {e1501524} (\bibinfo {year} {2016})}\BibitemShut {NoStop}%
\bibitem [{\citenamefont {{Parker}}\ \emph {et~al.}(2018)\citenamefont
  {{Parker}}, \citenamefont {{Morimoto}}, \citenamefont {{Orenstein}},\ and\
  \citenamefont {{Moore}}}]{Parker18}%
  \BibitemOpen
  \bibfield  {author} {\bibinfo {author} {\bibfnamefont {D.~E.}\ \bibnamefont
  {{Parker}}}, \bibinfo {author} {\bibfnamefont {T.}~\bibnamefont
  {{Morimoto}}}, \bibinfo {author} {\bibfnamefont {J.}~\bibnamefont
  {{Orenstein}}}, \ and\ \bibinfo {author} {\bibfnamefont {J.~E.}\ \bibnamefont
  {{Moore}}},\ }\href@noop {} {\bibfield  {journal} {\bibinfo  {journal} {ArXiv
  e-prints}\ ,\ \bibinfo {eid} {arXiv:1807.09285}} (\bibinfo {year} {2018})},\
  \Eprint {http://arxiv.org/abs/1807.09285} {arXiv:1807.09285} \BibitemShut
  {NoStop}%
\bibitem [{Sup(2019)}]{Supplementalphotoweyl}%
  \BibitemOpen
  \href@noop {} {\bibfield  {journal} {\bibinfo  {journal} {{Supplemental
  Information}}\ }\textbf {\bibinfo {volume} {X}},\ \bibinfo {pages} {X}
  (\bibinfo {year} {2019})}\BibitemShut {NoStop}%
\bibitem [{\citenamefont {Steiner}\ \emph {et~al.}(2017)\citenamefont
  {Steiner}, \citenamefont {Andreev},\ and\ \citenamefont {Pesin}}]{Steiner17}%
  \BibitemOpen
  \bibfield  {author} {\bibinfo {author} {\bibfnamefont {J.~F.}\ \bibnamefont
  {Steiner}}, \bibinfo {author} {\bibfnamefont {A.~V.}\ \bibnamefont
  {Andreev}}, \ and\ \bibinfo {author} {\bibfnamefont {D.~A.}\ \bibnamefont
  {Pesin}},\ }\href {\doibase 10.1103/PhysRevLett.119.036601} {\bibfield
  {journal} {\bibinfo  {journal} {Phys. Rev. Lett.}\ }\textbf {\bibinfo
  {volume} {119}},\ \bibinfo {pages} {036601} (\bibinfo {year}
  {2017})}\BibitemShut {NoStop}%
\bibitem [{\citenamefont {Deyo}\ \emph {et~al.}(2009)\citenamefont {Deyo},
  \citenamefont {Golub}, \citenamefont {Ivchenko},\ and\ \citenamefont
  {Spivak}}]{Deyo09}%
  \BibitemOpen
  \bibfield  {author} {\bibinfo {author} {\bibfnamefont {E.}~\bibnamefont
  {Deyo}}, \bibinfo {author} {\bibfnamefont {L.}~\bibnamefont {Golub}},
  \bibinfo {author} {\bibfnamefont {E.}~\bibnamefont {Ivchenko}}, \ and\
  \bibinfo {author} {\bibfnamefont {B.}~\bibnamefont {Spivak}},\ }\href@noop {}
  {\bibfield  {journal} {\bibinfo  {journal} {arXiv:0904.1917}\ } (\bibinfo
  {year} {2009})}\BibitemShut {NoStop}%
\bibitem [{\citenamefont {Moore}\ and\ \citenamefont
  {Orenstein}(2010)}]{MooreOrenstein}%
  \BibitemOpen
  \bibfield  {author} {\bibinfo {author} {\bibfnamefont {J.~E.}\ \bibnamefont
  {Moore}}\ and\ \bibinfo {author} {\bibfnamefont {J.}~\bibnamefont
  {Orenstein}},\ }\href {\doibase 10.1103/PhysRevLett.105.026805} {\bibfield
  {journal} {\bibinfo  {journal} {Phys. Rev. Lett.}\ }\textbf {\bibinfo
  {volume} {105}},\ \bibinfo {pages} {026805} (\bibinfo {year}
  {2010})}\BibitemShut {NoStop}%
\bibitem [{\citenamefont {Morimoto}\ \emph {et~al.}(2016)\citenamefont
  {Morimoto}, \citenamefont {Zhong}, \citenamefont {Orenstein},\ and\
  \citenamefont {Moore}}]{Morimoto16}%
  \BibitemOpen
  \bibfield  {author} {\bibinfo {author} {\bibfnamefont {T.}~\bibnamefont
  {Morimoto}}, \bibinfo {author} {\bibfnamefont {S.}~\bibnamefont {Zhong}},
  \bibinfo {author} {\bibfnamefont {J.}~\bibnamefont {Orenstein}}, \ and\
  \bibinfo {author} {\bibfnamefont {J.~E.}\ \bibnamefont {Moore}},\ }\href
  {\doibase 10.1103/PhysRevB.94.245121} {\bibfield  {journal} {\bibinfo
  {journal} {Phys. Rev. B}\ }\textbf {\bibinfo {volume} {94}},\ \bibinfo
  {pages} {245121} (\bibinfo {year} {2016})}\BibitemShut {NoStop}%
\bibitem [{\citenamefont {Ishizuka}\ \emph {et~al.}(2017)\citenamefont
  {Ishizuka}, \citenamefont {Hayata}, \citenamefont {Ueda},\ and\ \citenamefont
  {Nagaosa}}]{IHU16}%
  \BibitemOpen
  \bibfield  {author} {\bibinfo {author} {\bibfnamefont {H.}~\bibnamefont
  {Ishizuka}}, \bibinfo {author} {\bibfnamefont {T.}~\bibnamefont {Hayata}},
  \bibinfo {author} {\bibfnamefont {M.}~\bibnamefont {Ueda}}, \ and\ \bibinfo
  {author} {\bibfnamefont {N.}~\bibnamefont {Nagaosa}},\ }\href {\doibase
  10.1103/PhysRevB.95.245211} {\bibfield  {journal} {\bibinfo  {journal} {Phys.
  Rev. B}\ }\textbf {\bibinfo {volume} {95}},\ \bibinfo {pages} {245211}
  (\bibinfo {year} {2017})}\BibitemShut {NoStop}%
\bibitem [{\citenamefont {Rostami}\ and\ \citenamefont
  {Polini}(2017)}]{Rostami17}%
  \BibitemOpen
  \bibfield  {author} {\bibinfo {author} {\bibfnamefont {H.}~\bibnamefont
  {Rostami}}\ and\ \bibinfo {author} {\bibfnamefont {M.}~\bibnamefont
  {Polini}},\ }\href@noop {} {\bibfield  {journal} {\bibinfo  {journal}
  {arXiv:1705.09915}\ } (\bibinfo {year} {2017})}\BibitemShut {NoStop}%
\bibitem [{\citenamefont {Zhang}\ \emph
  {et~al.}(2018{\natexlab{b}})\citenamefont {Zhang}, \citenamefont {Sun},\ and\
  \citenamefont {Yan}}]{ZSY18}%
  \BibitemOpen
  \bibfield  {author} {\bibinfo {author} {\bibfnamefont {Y.}~\bibnamefont
  {Zhang}}, \bibinfo {author} {\bibfnamefont {Y.}~\bibnamefont {Sun}}, \ and\
  \bibinfo {author} {\bibfnamefont {B.}~\bibnamefont {Yan}},\ }\href {\doibase
  10.1103/PhysRevB.97.041101} {\bibfield  {journal} {\bibinfo  {journal} {Phys.
  Rev. B}\ }\textbf {\bibinfo {volume} {97}},\ \bibinfo {pages} {041101}
  (\bibinfo {year} {2018}{\natexlab{b}})}\BibitemShut {NoStop}%
\bibitem [{\citenamefont {Chan}\ \emph {et~al.}(2017)\citenamefont {Chan},
  \citenamefont {Lindner}, \citenamefont {Refael},\ and\ \citenamefont
  {Lee}}]{Chan16}%
  \BibitemOpen
  \bibfield  {author} {\bibinfo {author} {\bibfnamefont {C.-K.}\ \bibnamefont
  {Chan}}, \bibinfo {author} {\bibfnamefont {N.~H.}\ \bibnamefont {Lindner}},
  \bibinfo {author} {\bibfnamefont {G.}~\bibnamefont {Refael}}, \ and\ \bibinfo
  {author} {\bibfnamefont {P.~A.}\ \bibnamefont {Lee}},\ }\href {\doibase
  10.1103/PhysRevB.95.041104} {\bibfield  {journal} {\bibinfo  {journal} {Phys.
  Rev. B}\ }\textbf {\bibinfo {volume} {95}},\ \bibinfo {pages} {041104}
  (\bibinfo {year} {2017})}\BibitemShut {NoStop}%
\bibitem [{\citenamefont {K\"onig}\ \emph {et~al.}(2017)\citenamefont
  {K\"onig}, \citenamefont {Xie}, \citenamefont {Pesin},\ and\ \citenamefont
  {Levchenko}}]{Konig17}%
  \BibitemOpen
  \bibfield  {author} {\bibinfo {author} {\bibfnamefont {E.~J.}\ \bibnamefont
  {K\"onig}}, \bibinfo {author} {\bibfnamefont {H.-Y.}\ \bibnamefont {Xie}},
  \bibinfo {author} {\bibfnamefont {D.~A.}\ \bibnamefont {Pesin}}, \ and\
  \bibinfo {author} {\bibfnamefont {A.}~\bibnamefont {Levchenko}},\ }\href
  {\doibase 10.1103/PhysRevB.96.075123} {\bibfield  {journal} {\bibinfo
  {journal} {Phys. Rev. B}\ }\textbf {\bibinfo {volume} {96}},\ \bibinfo
  {pages} {075123} (\bibinfo {year} {2017})}\BibitemShut {NoStop}%
\bibitem [{\citenamefont {Golub}\ \emph {et~al.}(2017)\citenamefont {Golub},
  \citenamefont {Ivchenko},\ and\ \citenamefont {Spivak}}]{Golub17}%
  \BibitemOpen
  \bibfield  {author} {\bibinfo {author} {\bibfnamefont {L.}~\bibnamefont
  {Golub}}, \bibinfo {author} {\bibfnamefont {E.~L.}\ \bibnamefont {Ivchenko}},
  \ and\ \bibinfo {author} {\bibfnamefont {B.}~\bibnamefont {Spivak}},\
  }\href@noop {} {\bibfield  {journal} {\bibinfo  {journal} {JETP Letters}\
  }\textbf {\bibinfo {volume} {105}},\ \bibinfo {pages} {782} (\bibinfo {year}
  {2017})}\BibitemShut {NoStop}%
\bibitem [{\citenamefont {Ma\~nes}(2012)}]{manes2012}%
  \BibitemOpen
  \bibfield  {author} {\bibinfo {author} {\bibfnamefont {J.~L.}\ \bibnamefont
  {Ma\~nes}},\ }\href {\doibase 10.1103/PhysRevB.85.155118} {\bibfield
  {journal} {\bibinfo  {journal} {Phys. Rev. B}\ }\textbf {\bibinfo {volume}
  {85}},\ \bibinfo {pages} {155118} (\bibinfo {year} {2012})}\BibitemShut
  {NoStop}%
\bibitem [{\citenamefont {Bradlyn}\ \emph {et~al.}(2016)\citenamefont
  {Bradlyn}, \citenamefont {Cano}, \citenamefont {Wang}, \citenamefont
  {Vergniory}, \citenamefont {Felser}, \citenamefont {Cava},\ and\
  \citenamefont {Bernevig}}]{BradlynEA17}%
  \BibitemOpen
  \bibfield  {author} {\bibinfo {author} {\bibfnamefont {B.}~\bibnamefont
  {Bradlyn}}, \bibinfo {author} {\bibfnamefont {J.}~\bibnamefont {Cano}},
  \bibinfo {author} {\bibfnamefont {Z.}~\bibnamefont {Wang}}, \bibinfo {author}
  {\bibfnamefont {M.~G.}\ \bibnamefont {Vergniory}}, \bibinfo {author}
  {\bibfnamefont {C.}~\bibnamefont {Felser}}, \bibinfo {author} {\bibfnamefont
  {R.~J.}\ \bibnamefont {Cava}}, \ and\ \bibinfo {author} {\bibfnamefont
  {B.~A.}\ \bibnamefont {Bernevig}},\ }\href {\doibase 10.1126/science.aaf5037}
  {\bibfield  {journal} {\bibinfo  {journal} {Science}\ }\textbf {\bibinfo
  {volume} {353}},\ \bibinfo {pages} {6299} (\bibinfo {year}
  {2016})}\BibitemShut {NoStop}%
\bibitem [{\citenamefont {Chang}\ \emph {et~al.}(2017)\citenamefont {Chang},
  \citenamefont {Xu}, \citenamefont {Wieder}, \citenamefont {Sanchez},
  \citenamefont {Huang}, \citenamefont {Belopolski}, \citenamefont {Chang},
  \citenamefont {Zhang}, \citenamefont {Bansil}, \citenamefont {Lin},\ and\
  \citenamefont {Hasan}}]{ChangEA17}%
  \BibitemOpen
  \bibfield  {author} {\bibinfo {author} {\bibfnamefont {G.}~\bibnamefont
  {Chang}}, \bibinfo {author} {\bibfnamefont {S.-Y.}\ \bibnamefont {Xu}},
  \bibinfo {author} {\bibfnamefont {B.~J.}\ \bibnamefont {Wieder}}, \bibinfo
  {author} {\bibfnamefont {D.~S.}\ \bibnamefont {Sanchez}}, \bibinfo {author}
  {\bibfnamefont {S.-M.}\ \bibnamefont {Huang}}, \bibinfo {author}
  {\bibfnamefont {I.}~\bibnamefont {Belopolski}}, \bibinfo {author}
  {\bibfnamefont {T.-R.}\ \bibnamefont {Chang}}, \bibinfo {author}
  {\bibfnamefont {S.}~\bibnamefont {Zhang}}, \bibinfo {author} {\bibfnamefont
  {A.}~\bibnamefont {Bansil}}, \bibinfo {author} {\bibfnamefont
  {H.}~\bibnamefont {Lin}}, \ and\ \bibinfo {author} {\bibfnamefont {M.~Z.}\
  \bibnamefont {Hasan}},\ }\href {\doibase 10.1103/PhysRevLett.119.206401}
  {\bibfield  {journal} {\bibinfo  {journal} {Phys. Rev. Lett.}\ }\textbf
  {\bibinfo {volume} {119}},\ \bibinfo {pages} {206401} (\bibinfo {year}
  {2017})}\BibitemShut {NoStop}%
\bibitem [{\citenamefont {Tang}\ \emph {et~al.}(2017)\citenamefont {Tang},
  \citenamefont {Zhou},\ and\ \citenamefont {Zhang}}]{TangEA17}%
  \BibitemOpen
  \bibfield  {author} {\bibinfo {author} {\bibfnamefont {P.}~\bibnamefont
  {Tang}}, \bibinfo {author} {\bibfnamefont {Q.}~\bibnamefont {Zhou}}, \ and\
  \bibinfo {author} {\bibfnamefont {S.-C.}\ \bibnamefont {Zhang}},\ }\href
  {\doibase 10.1103/PhysRevLett.119.206402} {\bibfield  {journal} {\bibinfo
  {journal} {Phys. Rev. Lett.}\ }\textbf {\bibinfo {volume} {119}},\ \bibinfo
  {pages} {206402} (\bibinfo {year} {2017})}\BibitemShut {NoStop}%
\bibitem [{\citenamefont {Bouhon}\ and\ \citenamefont
  {Black-Schaffer}(2017)}]{Bouhon2017}%
  \BibitemOpen
  \bibfield  {author} {\bibinfo {author} {\bibfnamefont {A.}~\bibnamefont
  {Bouhon}}\ and\ \bibinfo {author} {\bibfnamefont {A.~M.}\ \bibnamefont
  {Black-Schaffer}},\ }\href {\doibase 10.1103/PhysRevB.95.241101} {\bibfield
  {journal} {\bibinfo  {journal} {Phys. Rev. B}\ }\textbf {\bibinfo {volume}
  {95}},\ \bibinfo {pages} {241101} (\bibinfo {year} {2017})}\BibitemShut
  {NoStop}%
\bibitem [{\citenamefont {Takane}\ \emph {et~al.}(2019)\citenamefont {Takane},
  \citenamefont {Wang}, \citenamefont {Souma}, \citenamefont {Nakayama},
  \citenamefont {Nakamura}, \citenamefont {Oinuma}, \citenamefont {Nakata},
  \citenamefont {Iwasawa}, \citenamefont {Cacho}, \citenamefont {Kim},
  \citenamefont {Horiba}, \citenamefont {Kumigashira}, \citenamefont
  {Takahashi}, \citenamefont {Ando},\ and\ \citenamefont {Sato}}]{Takane2019}%
  \BibitemOpen
  \bibfield  {author} {\bibinfo {author} {\bibfnamefont {D.}~\bibnamefont
  {Takane}}, \bibinfo {author} {\bibfnamefont {Z.}~\bibnamefont {Wang}},
  \bibinfo {author} {\bibfnamefont {S.}~\bibnamefont {Souma}}, \bibinfo
  {author} {\bibfnamefont {K.}~\bibnamefont {Nakayama}}, \bibinfo {author}
  {\bibfnamefont {T.}~\bibnamefont {Nakamura}}, \bibinfo {author}
  {\bibfnamefont {H.}~\bibnamefont {Oinuma}}, \bibinfo {author} {\bibfnamefont
  {Y.}~\bibnamefont {Nakata}}, \bibinfo {author} {\bibfnamefont
  {H.}~\bibnamefont {Iwasawa}}, \bibinfo {author} {\bibfnamefont
  {C.}~\bibnamefont {Cacho}}, \bibinfo {author} {\bibfnamefont
  {T.}~\bibnamefont {Kim}}, \bibinfo {author} {\bibfnamefont {K.}~\bibnamefont
  {Horiba}}, \bibinfo {author} {\bibfnamefont {H.}~\bibnamefont {Kumigashira}},
  \bibinfo {author} {\bibfnamefont {T.}~\bibnamefont {Takahashi}}, \bibinfo
  {author} {\bibfnamefont {Y.}~\bibnamefont {Ando}}, \ and\ \bibinfo {author}
  {\bibfnamefont {T.}~\bibnamefont {Sato}},\ }\href {\doibase
  10.1103/PhysRevLett.122.076402} {\bibfield  {journal} {\bibinfo  {journal}
  {Phys. Rev. Lett.}\ }\textbf {\bibinfo {volume} {122}},\ \bibinfo {pages}
  {076402} (\bibinfo {year} {2019})}\BibitemShut {NoStop}%
\bibitem [{\citenamefont {Rao}\ \emph {et~al.}(2019)\citenamefont {Rao},
  \citenamefont {Li}, \citenamefont {Zhang}, \citenamefont {Tian},
  \citenamefont {Li}, \citenamefont {Fu}, \citenamefont {Tang}, \citenamefont
  {Wang}, \citenamefont {Li}, \citenamefont {Fan}, \citenamefont {Li},
  \citenamefont {Huang}, \citenamefont {Liu}, \citenamefont {Long},
  \citenamefont {Fang}, \citenamefont {Weng}, \citenamefont {Shi},
  \citenamefont {Lei}, \citenamefont {Sun}, \citenamefont {Qian},\ and\
  \citenamefont {Ding}}]{Rao:2019uw}%
  \BibitemOpen
  \bibfield  {author} {\bibinfo {author} {\bibfnamefont {Z.}~\bibnamefont
  {Rao}}, \bibinfo {author} {\bibfnamefont {H.}~\bibnamefont {Li}}, \bibinfo
  {author} {\bibfnamefont {T.}~\bibnamefont {Zhang}}, \bibinfo {author}
  {\bibfnamefont {S.}~\bibnamefont {Tian}}, \bibinfo {author} {\bibfnamefont
  {C.}~\bibnamefont {Li}}, \bibinfo {author} {\bibfnamefont {B.}~\bibnamefont
  {Fu}}, \bibinfo {author} {\bibfnamefont {C.}~\bibnamefont {Tang}}, \bibinfo
  {author} {\bibfnamefont {L.}~\bibnamefont {Wang}}, \bibinfo {author}
  {\bibfnamefont {Z.}~\bibnamefont {Li}}, \bibinfo {author} {\bibfnamefont
  {W.}~\bibnamefont {Fan}}, \bibinfo {author} {\bibfnamefont {J.}~\bibnamefont
  {Li}}, \bibinfo {author} {\bibfnamefont {Y.}~\bibnamefont {Huang}}, \bibinfo
  {author} {\bibfnamefont {Z.}~\bibnamefont {Liu}}, \bibinfo {author}
  {\bibfnamefont {Y.}~\bibnamefont {Long}}, \bibinfo {author} {\bibfnamefont
  {C.}~\bibnamefont {Fang}}, \bibinfo {author} {\bibfnamefont {H.}~\bibnamefont
  {Weng}}, \bibinfo {author} {\bibfnamefont {Y.}~\bibnamefont {Shi}}, \bibinfo
  {author} {\bibfnamefont {H.}~\bibnamefont {Lei}}, \bibinfo {author}
  {\bibfnamefont {Y.}~\bibnamefont {Sun}}, \bibinfo {author} {\bibfnamefont
  {T.}~\bibnamefont {Qian}}, \ and\ \bibinfo {author} {\bibfnamefont
  {H.}~\bibnamefont {Ding}},\ }\href
  {https://doi.org/10.1038/s41586-019-1031-8} {\bibfield  {journal} {\bibinfo
  {journal} {Nature}\ }\textbf {\bibinfo {volume} {567}},\ \bibinfo {pages}
  {496} (\bibinfo {year} {2019})}\BibitemShut {NoStop}%
\bibitem [{\citenamefont {Sanchez}\ \emph {et~al.}(2019)\citenamefont
  {Sanchez}, \citenamefont {Belopolski}, \citenamefont {Cochran}, \citenamefont
  {Xu}, \citenamefont {Yin}, \citenamefont {Chang}, \citenamefont {Xie},
  \citenamefont {Manna}, \citenamefont {S{\"u}{\ss}}, \citenamefont {Huang},
  \citenamefont {Alidoust}, \citenamefont {Multer}, \citenamefont {Zhang},
  \citenamefont {Shumiya}, \citenamefont {Wang}, \citenamefont {Wang},
  \citenamefont {Chang}, \citenamefont {Felser}, \citenamefont {Xu},
  \citenamefont {Jia}, \citenamefont {Lin},\ and\ \citenamefont
  {Hasan}}]{Sanchez:2019wl}%
  \BibitemOpen
  \bibfield  {author} {\bibinfo {author} {\bibfnamefont {D.~S.}\ \bibnamefont
  {Sanchez}}, \bibinfo {author} {\bibfnamefont {I.}~\bibnamefont {Belopolski}},
  \bibinfo {author} {\bibfnamefont {T.~A.}\ \bibnamefont {Cochran}}, \bibinfo
  {author} {\bibfnamefont {X.}~\bibnamefont {Xu}}, \bibinfo {author}
  {\bibfnamefont {J.-X.}\ \bibnamefont {Yin}}, \bibinfo {author} {\bibfnamefont
  {G.}~\bibnamefont {Chang}}, \bibinfo {author} {\bibfnamefont
  {W.}~\bibnamefont {Xie}}, \bibinfo {author} {\bibfnamefont {K.}~\bibnamefont
  {Manna}}, \bibinfo {author} {\bibfnamefont {V.}~\bibnamefont {S{\"u}{\ss}}},
  \bibinfo {author} {\bibfnamefont {C.-Y.}\ \bibnamefont {Huang}}, \bibinfo
  {author} {\bibfnamefont {N.}~\bibnamefont {Alidoust}}, \bibinfo {author}
  {\bibfnamefont {D.}~\bibnamefont {Multer}}, \bibinfo {author} {\bibfnamefont
  {S.~S.}\ \bibnamefont {Zhang}}, \bibinfo {author} {\bibfnamefont
  {N.}~\bibnamefont {Shumiya}}, \bibinfo {author} {\bibfnamefont
  {X.}~\bibnamefont {Wang}}, \bibinfo {author} {\bibfnamefont {G.-Q.}\
  \bibnamefont {Wang}}, \bibinfo {author} {\bibfnamefont {T.-R.}\ \bibnamefont
  {Chang}}, \bibinfo {author} {\bibfnamefont {C.}~\bibnamefont {Felser}},
  \bibinfo {author} {\bibfnamefont {S.-Y.}\ \bibnamefont {Xu}}, \bibinfo
  {author} {\bibfnamefont {S.}~\bibnamefont {Jia}}, \bibinfo {author}
  {\bibfnamefont {H.}~\bibnamefont {Lin}}, \ and\ \bibinfo {author}
  {\bibfnamefont {M.~Z.}\ \bibnamefont {Hasan}},\ }\href
  {https://doi.org/10.1038/s41586-019-1037-2} {\bibfield  {journal} {\bibinfo
  {journal} {Nature}\ }\textbf {\bibinfo {volume} {567}},\ \bibinfo {pages}
  {500} (\bibinfo {year} {2019})}\BibitemShut {NoStop}%
\bibitem [{\citenamefont {Schr{\"o}ter}\ \emph {et~al.}(2019)\citenamefont
  {Schr{\"o}ter}, \citenamefont {Pei}, \citenamefont {Vergniory}, \citenamefont
  {Sun}, \citenamefont {Manna}, \citenamefont {de~Juan}, \citenamefont
  {Krieger}, \citenamefont {S{\"u}{\ss}}, \citenamefont {Schmidt},
  \citenamefont {Dudin}, \citenamefont {Bradlyn}, \citenamefont {Kim},
  \citenamefont {Schmitt}, \citenamefont {Cacho}, \citenamefont {Felser},
  \citenamefont {Strocov},\ and\ \citenamefont {Chen}}]{Schroter:2019kf}%
  \BibitemOpen
  \bibfield  {author} {\bibinfo {author} {\bibfnamefont {N.~B.~M.}\
  \bibnamefont {Schr{\"o}ter}}, \bibinfo {author} {\bibfnamefont
  {D.}~\bibnamefont {Pei}}, \bibinfo {author} {\bibfnamefont {M.~G.}\
  \bibnamefont {Vergniory}}, \bibinfo {author} {\bibfnamefont {Y.}~\bibnamefont
  {Sun}}, \bibinfo {author} {\bibfnamefont {K.}~\bibnamefont {Manna}}, \bibinfo
  {author} {\bibfnamefont {F.}~\bibnamefont {de~Juan}}, \bibinfo {author}
  {\bibfnamefont {J.~A.}\ \bibnamefont {Krieger}}, \bibinfo {author}
  {\bibfnamefont {V.}~\bibnamefont {S{\"u}{\ss}}}, \bibinfo {author}
  {\bibfnamefont {M.}~\bibnamefont {Schmidt}}, \bibinfo {author} {\bibfnamefont
  {P.}~\bibnamefont {Dudin}}, \bibinfo {author} {\bibfnamefont
  {B.}~\bibnamefont {Bradlyn}}, \bibinfo {author} {\bibfnamefont {T.~K.}\
  \bibnamefont {Kim}}, \bibinfo {author} {\bibfnamefont {T.}~\bibnamefont
  {Schmitt}}, \bibinfo {author} {\bibfnamefont {C.}~\bibnamefont {Cacho}},
  \bibinfo {author} {\bibfnamefont {C.}~\bibnamefont {Felser}}, \bibinfo
  {author} {\bibfnamefont {V.~N.}\ \bibnamefont {Strocov}}, \ and\ \bibinfo
  {author} {\bibfnamefont {Y.}~\bibnamefont {Chen}},\ }\href {\doibase
  10.1038/s41567-019-0511-y} {\bibfield  {journal} {\bibinfo  {journal} {Nature
  Physics}\ }\textbf {\bibinfo {volume} {353}},\ \bibinfo {pages} {1} (\bibinfo
  {year} {2019})}\BibitemShut {NoStop}%
\bibitem [{\citenamefont {Weng}\ \emph {et~al.}(2015)\citenamefont {Weng},
  \citenamefont {Fang}, \citenamefont {Fang}, \citenamefont {Bernevig},\ and\
  \citenamefont {Dai}}]{Weng2015}%
  \BibitemOpen
  \bibfield  {author} {\bibinfo {author} {\bibfnamefont {H.}~\bibnamefont
  {Weng}}, \bibinfo {author} {\bibfnamefont {C.}~\bibnamefont {Fang}}, \bibinfo
  {author} {\bibfnamefont {Z.}~\bibnamefont {Fang}}, \bibinfo {author}
  {\bibfnamefont {B.~A.}\ \bibnamefont {Bernevig}}, \ and\ \bibinfo {author}
  {\bibfnamefont {X.}~\bibnamefont {Dai}},\ }\href {\doibase
  10.1103/PhysRevX.5.011029} {\bibfield  {journal} {\bibinfo  {journal} {Phys.
  Rev. X}\ }\textbf {\bibinfo {volume} {5}},\ \bibinfo {pages} {011029}
  (\bibinfo {year} {2015})}\BibitemShut {NoStop}%
\bibitem [{\citenamefont {Huang}\ \emph {et~al.}(2015)\citenamefont {Huang},
  \citenamefont {Xu}, \citenamefont {Belopolski}, \citenamefont {Lee},
  \citenamefont {Chang}, \citenamefont {Wang}, \citenamefont {Alidoust},
  \citenamefont {Bian}, \citenamefont {Neupane}, \citenamefont {Zhang},
  \citenamefont {Jia}, \citenamefont {Bansil}, \citenamefont {Lin},\ and\
  \citenamefont {Hasan}}]{Huang:vn}%
  \BibitemOpen
  \bibfield  {author} {\bibinfo {author} {\bibfnamefont {S.-M.}\ \bibnamefont
  {Huang}}, \bibinfo {author} {\bibfnamefont {S.-Y.}\ \bibnamefont {Xu}},
  \bibinfo {author} {\bibfnamefont {I.}~\bibnamefont {Belopolski}}, \bibinfo
  {author} {\bibfnamefont {C.-C.}\ \bibnamefont {Lee}}, \bibinfo {author}
  {\bibfnamefont {G.}~\bibnamefont {Chang}}, \bibinfo {author} {\bibfnamefont
  {B.}~\bibnamefont {Wang}}, \bibinfo {author} {\bibfnamefont {N.}~\bibnamefont
  {Alidoust}}, \bibinfo {author} {\bibfnamefont {G.}~\bibnamefont {Bian}},
  \bibinfo {author} {\bibfnamefont {M.}~\bibnamefont {Neupane}}, \bibinfo
  {author} {\bibfnamefont {C.}~\bibnamefont {Zhang}}, \bibinfo {author}
  {\bibfnamefont {S.}~\bibnamefont {Jia}}, \bibinfo {author} {\bibfnamefont
  {A.}~\bibnamefont {Bansil}}, \bibinfo {author} {\bibfnamefont
  {H.}~\bibnamefont {Lin}}, \ and\ \bibinfo {author} {\bibfnamefont {M.~Z.}\
  \bibnamefont {Hasan}},\ }\href {https://doi.org/10.1038/ncomms8373}
  {\bibfield  {journal} {\bibinfo  {journal} {Nature Communications}\ }\textbf
  {\bibinfo {volume} {6}},\ \bibinfo {pages} {7373} (\bibinfo {year}
  {2015})}\BibitemShut {NoStop}%
\bibitem [{\citenamefont {Lv}\ \emph {et~al.}(2015{\natexlab{a}})\citenamefont
  {Lv}, \citenamefont {Xu}, \citenamefont {Weng}, \citenamefont {Ma},
  \citenamefont {Richard}, \citenamefont {Huang}, \citenamefont {Zhao},
  \citenamefont {Chen}, \citenamefont {Matt}, \citenamefont {Bisti},
  \citenamefont {Strocov}, \citenamefont {Mesot}, \citenamefont {Fang},
  \citenamefont {Dai}, \citenamefont {Qian}, \citenamefont {Shi},\ and\
  \citenamefont {Ding}}]{LvXu2015}%
  \BibitemOpen
  \bibfield  {author} {\bibinfo {author} {\bibfnamefont {B.~Q.}\ \bibnamefont
  {Lv}}, \bibinfo {author} {\bibfnamefont {N.}~\bibnamefont {Xu}}, \bibinfo
  {author} {\bibfnamefont {H.~M.}\ \bibnamefont {Weng}}, \bibinfo {author}
  {\bibfnamefont {J.~Z.}\ \bibnamefont {Ma}}, \bibinfo {author} {\bibfnamefont
  {P.}~\bibnamefont {Richard}}, \bibinfo {author} {\bibfnamefont {X.~C.}\
  \bibnamefont {Huang}}, \bibinfo {author} {\bibfnamefont {L.~X.}\ \bibnamefont
  {Zhao}}, \bibinfo {author} {\bibfnamefont {G.~F.}\ \bibnamefont {Chen}},
  \bibinfo {author} {\bibfnamefont {C.~E.}\ \bibnamefont {Matt}}, \bibinfo
  {author} {\bibfnamefont {F.}~\bibnamefont {Bisti}}, \bibinfo {author}
  {\bibfnamefont {V.~N.}\ \bibnamefont {Strocov}}, \bibinfo {author}
  {\bibfnamefont {J.}~\bibnamefont {Mesot}}, \bibinfo {author} {\bibfnamefont
  {Z.}~\bibnamefont {Fang}}, \bibinfo {author} {\bibfnamefont {X.}~\bibnamefont
  {Dai}}, \bibinfo {author} {\bibfnamefont {T.}~\bibnamefont {Qian}}, \bibinfo
  {author} {\bibfnamefont {M.}~\bibnamefont {Shi}}, \ and\ \bibinfo {author}
  {\bibfnamefont {H.}~\bibnamefont {Ding}},\ }\href {\doibase
  10.1038/nphys3426} {\bibfield  {journal} {\bibinfo  {journal} {Nature
  Physics}\ }\textbf {\bibinfo {volume} {11}},\ \bibinfo {pages} {724}
  (\bibinfo {year} {2015}{\natexlab{a}})}\BibitemShut {NoStop}%
\bibitem [{\citenamefont {Xu}\ \emph {et~al.}(2015)\citenamefont {Xu},
  \citenamefont {Belopolski}, \citenamefont {Alidoust}, \citenamefont
  {Neupane}, \citenamefont {Bian}, \citenamefont {Zhang}, \citenamefont
  {Sankar}, \citenamefont {Chang}, \citenamefont {Yuan}, \citenamefont {Lee},
  \citenamefont {Huang}, \citenamefont {Zheng}, \citenamefont {Ma},
  \citenamefont {Sanchez}, \citenamefont {Wang}, \citenamefont {Bansil},
  \citenamefont {Chou}, \citenamefont {Shibayev}, \citenamefont {Lin},
  \citenamefont {Jia},\ and\ \citenamefont {Hasan}}]{Xu613}%
  \BibitemOpen
  \bibfield  {author} {\bibinfo {author} {\bibfnamefont {S.-Y.}\ \bibnamefont
  {Xu}}, \bibinfo {author} {\bibfnamefont {I.}~\bibnamefont {Belopolski}},
  \bibinfo {author} {\bibfnamefont {N.}~\bibnamefont {Alidoust}}, \bibinfo
  {author} {\bibfnamefont {M.}~\bibnamefont {Neupane}}, \bibinfo {author}
  {\bibfnamefont {G.}~\bibnamefont {Bian}}, \bibinfo {author} {\bibfnamefont
  {C.}~\bibnamefont {Zhang}}, \bibinfo {author} {\bibfnamefont
  {R.}~\bibnamefont {Sankar}}, \bibinfo {author} {\bibfnamefont
  {G.}~\bibnamefont {Chang}}, \bibinfo {author} {\bibfnamefont
  {Z.}~\bibnamefont {Yuan}}, \bibinfo {author} {\bibfnamefont {C.-C.}\
  \bibnamefont {Lee}}, \bibinfo {author} {\bibfnamefont {S.-M.}\ \bibnamefont
  {Huang}}, \bibinfo {author} {\bibfnamefont {H.}~\bibnamefont {Zheng}},
  \bibinfo {author} {\bibfnamefont {J.}~\bibnamefont {Ma}}, \bibinfo {author}
  {\bibfnamefont {D.~S.}\ \bibnamefont {Sanchez}}, \bibinfo {author}
  {\bibfnamefont {B.}~\bibnamefont {Wang}}, \bibinfo {author} {\bibfnamefont
  {A.}~\bibnamefont {Bansil}}, \bibinfo {author} {\bibfnamefont
  {F.}~\bibnamefont {Chou}}, \bibinfo {author} {\bibfnamefont {P.~P.}\
  \bibnamefont {Shibayev}}, \bibinfo {author} {\bibfnamefont {H.}~\bibnamefont
  {Lin}}, \bibinfo {author} {\bibfnamefont {S.}~\bibnamefont {Jia}}, \ and\
  \bibinfo {author} {\bibfnamefont {M.~Z.}\ \bibnamefont {Hasan}},\ }\href
  {\doibase 10.1126/science.aaa9297} {\bibfield  {journal} {\bibinfo  {journal}
  {Science}\ }\textbf {\bibinfo {volume} {349}},\ \bibinfo {pages} {613}
  (\bibinfo {year} {2015})}\BibitemShut {NoStop}%
\bibitem [{\citenamefont {Yang}\ \emph {et~al.}(2015)\citenamefont {Yang},
  \citenamefont {Liu}, \citenamefont {Sun}, \citenamefont {Peng}, \citenamefont
  {Yang}, \citenamefont {Zhang}, \citenamefont {Zhou}, \citenamefont {Zhang},
  \citenamefont {Guo}, \citenamefont {Rahn}, \citenamefont {Prabhakaran},
  \citenamefont {Hussain}, \citenamefont {Mo}, \citenamefont {Felser},
  \citenamefont {Yan},\ and\ \citenamefont {Chen}}]{Yang:2015ev}%
  \BibitemOpen
  \bibfield  {author} {\bibinfo {author} {\bibfnamefont {L.~X.}\ \bibnamefont
  {Yang}}, \bibinfo {author} {\bibfnamefont {Z.~K.}\ \bibnamefont {Liu}},
  \bibinfo {author} {\bibfnamefont {Y.}~\bibnamefont {Sun}}, \bibinfo {author}
  {\bibfnamefont {H.}~\bibnamefont {Peng}}, \bibinfo {author} {\bibfnamefont
  {H.~F.}\ \bibnamefont {Yang}}, \bibinfo {author} {\bibfnamefont
  {T.}~\bibnamefont {Zhang}}, \bibinfo {author} {\bibfnamefont
  {B.}~\bibnamefont {Zhou}}, \bibinfo {author} {\bibfnamefont {Y.}~\bibnamefont
  {Zhang}}, \bibinfo {author} {\bibfnamefont {Y.~F.}\ \bibnamefont {Guo}},
  \bibinfo {author} {\bibfnamefont {M.}~\bibnamefont {Rahn}}, \bibinfo {author}
  {\bibfnamefont {D.}~\bibnamefont {Prabhakaran}}, \bibinfo {author}
  {\bibfnamefont {Z.}~\bibnamefont {Hussain}}, \bibinfo {author} {\bibfnamefont
  {S.~K.}\ \bibnamefont {Mo}}, \bibinfo {author} {\bibfnamefont
  {C.}~\bibnamefont {Felser}}, \bibinfo {author} {\bibfnamefont
  {B.}~\bibnamefont {Yan}}, \ and\ \bibinfo {author} {\bibfnamefont {Y.~L.}\
  \bibnamefont {Chen}},\ }\href {\doibase 10.1038/nphys3425} {\bibfield
  {journal} {\bibinfo  {journal} {Nature Physics}\ }\textbf {\bibinfo {volume}
  {11}},\ \bibinfo {pages} {728} (\bibinfo {year} {2015})}\BibitemShut
  {NoStop}%
\bibitem [{\citenamefont {Leppenen}\ \emph {et~al.}(2019)\citenamefont
  {Leppenen}, \citenamefont {Ivchenko},\ and\ \citenamefont {Golub}}]{Golub19}%
  \BibitemOpen
  \bibfield  {author} {\bibinfo {author} {\bibfnamefont {N.}~\bibnamefont
  {Leppenen}}, \bibinfo {author} {\bibfnamefont {E.}~\bibnamefont {Ivchenko}},
  \ and\ \bibinfo {author} {\bibfnamefont {L.}~\bibnamefont {Golub}},\
  }\href@noop {} {\bibfield  {journal} {\bibinfo  {journal} {arXiv:1905.12273}\
  } (\bibinfo {year} {2019})}\BibitemShut {NoStop}%
\bibitem [{\citenamefont {Buckeridge}\ \emph {et~al.}(2016)\citenamefont
  {Buckeridge}, \citenamefont {Jevdokimovs}, \citenamefont {Catlow},\ and\
  \citenamefont {Sokol}}]{Buckeridge:2016fb}%
  \BibitemOpen
  \bibfield  {author} {\bibinfo {author} {\bibfnamefont {J.}~\bibnamefont
  {Buckeridge}}, \bibinfo {author} {\bibfnamefont {D.}~\bibnamefont
  {Jevdokimovs}}, \bibinfo {author} {\bibfnamefont {C.~R.~A.}\ \bibnamefont
  {Catlow}}, \ and\ \bibinfo {author} {\bibfnamefont {A.~A.}\ \bibnamefont
  {Sokol}},\ }\href {\doibase 10.1103/PhysRevB.93.125205} {\bibfield  {journal}
  {\bibinfo  {journal} {Physical Review B}\ }\textbf {\bibinfo {volume} {93}},\
  \bibinfo {pages} {125205} (\bibinfo {year} {2016})}\BibitemShut {NoStop}%
\bibitem [{\citenamefont {Belinicher}\ and\ \citenamefont
  {Sturman}(1980)}]{BS80}%
  \BibitemOpen
  \bibfield  {author} {\bibinfo {author} {\bibfnamefont {V.~I.}\ \bibnamefont
  {Belinicher}}\ and\ \bibinfo {author} {\bibfnamefont {B.~I.}\ \bibnamefont
  {Sturman}},\ }\href@noop {} {\bibfield  {journal} {\bibinfo  {journal} {Sov.
  Phys. Usp.}\ }\textbf {\bibinfo {volume} {23}},\ \bibinfo {pages} {199}
  (\bibinfo {year} {1980})}\BibitemShut {NoStop}%
\bibitem [{\citenamefont {Isobe}\ \emph {et~al.}(2018)\citenamefont {Isobe},
  \citenamefont {Xu},\ and\ \citenamefont {Fu}}]{Isobe18}%
  \BibitemOpen
  \bibfield  {author} {\bibinfo {author} {\bibfnamefont {H.}~\bibnamefont
  {Isobe}}, \bibinfo {author} {\bibfnamefont {S.-Y.}\ \bibnamefont {Xu}}, \
  and\ \bibinfo {author} {\bibfnamefont {L.}~\bibnamefont {Fu}},\ }\href@noop
  {} {\bibfield  {journal} {\bibinfo  {journal} {arXiv:1812.08162}\ } (\bibinfo
  {year} {2018})}\BibitemShut {NoStop}%
\bibitem [{\citenamefont {Golub}\ and\ \citenamefont
  {Ivchenko}(2018)}]{Golub18}%
  \BibitemOpen
  \bibfield  {author} {\bibinfo {author} {\bibfnamefont {L.~E.}\ \bibnamefont
  {Golub}}\ and\ \bibinfo {author} {\bibfnamefont {E.~L.}\ \bibnamefont
  {Ivchenko}},\ }\href {\doibase 10.1103/PhysRevB.98.075305} {\bibfield
  {journal} {\bibinfo  {journal} {Phys. Rev. B}\ }\textbf {\bibinfo {volume}
  {98}},\ \bibinfo {pages} {075305} (\bibinfo {year} {2018})}\BibitemShut
  {NoStop}%
\bibitem [{\citenamefont {Nandy}\ and\ \citenamefont
  {Sodemann}(2019)}]{Sodemann19}%
  \BibitemOpen
  \bibfield  {author} {\bibinfo {author} {\bibfnamefont {S.}~\bibnamefont
  {Nandy}}\ and\ \bibinfo {author} {\bibfnamefont {I.}~\bibnamefont
  {Sodemann}},\ }\href@noop {} {\bibfield  {journal} {\bibinfo  {journal}
  {arXiv:1901.04467}\ } (\bibinfo {year} {2019})}\BibitemShut {NoStop}%
\bibitem [{\citenamefont {Matsyshyn}\ and\ \citenamefont
  {Sodemann}(2019)}]{intiack}%
  \BibitemOpen
  \bibfield  {author} {\bibinfo {author} {\bibfnamefont {O.}~\bibnamefont
  {Matsyshyn}}\ and\ \bibinfo {author} {\bibfnamefont {I.}~\bibnamefont
  {Sodemann}},\ }\href@noop {} {\bibfield  {journal} {\bibinfo  {journal}
  {private communication and arXiv: 1907.XXXXX}\ } (\bibinfo {year}
  {2019})}\BibitemShut {NoStop}%
\bibitem [{\citenamefont {Hipolito}\ \emph {et~al.}(2016)\citenamefont
  {Hipolito}, \citenamefont {Pedersen},\ and\ \citenamefont
  {Pereira}}]{Hipolito16}%
  \BibitemOpen
  \bibfield  {author} {\bibinfo {author} {\bibfnamefont {F.}~\bibnamefont
  {Hipolito}}, \bibinfo {author} {\bibfnamefont {T.~G.}\ \bibnamefont
  {Pedersen}}, \ and\ \bibinfo {author} {\bibfnamefont {V.~M.}\ \bibnamefont
  {Pereira}},\ }\href {\doibase 10.1103/PhysRevB.94.045434} {\bibfield
  {journal} {\bibinfo  {journal} {Phys. Rev. B}\ }\textbf {\bibinfo {volume}
  {94}},\ \bibinfo {pages} {045434} (\bibinfo {year} {2016})}\BibitemShut
  {NoStop}%
\bibitem [{\citenamefont {Culcer}\ \emph {et~al.}(2017)\citenamefont {Culcer},
  \citenamefont {Sekine},\ and\ \citenamefont {MacDonald}}]{Culcer17}%
  \BibitemOpen
  \bibfield  {author} {\bibinfo {author} {\bibfnamefont {D.}~\bibnamefont
  {Culcer}}, \bibinfo {author} {\bibfnamefont {A.}~\bibnamefont {Sekine}}, \
  and\ \bibinfo {author} {\bibfnamefont {A.~H.}\ \bibnamefont {MacDonald}},\
  }\href {\doibase 10.1103/PhysRevB.96.035106} {\bibfield  {journal} {\bibinfo
  {journal} {Phys. Rev. B}\ }\textbf {\bibinfo {volume} {96}},\ \bibinfo
  {pages} {035106} (\bibinfo {year} {2017})}\BibitemShut {NoStop}%
\bibitem [{\citenamefont {Gao}\ \emph {et~al.}(2014)\citenamefont {Gao},
  \citenamefont {Yang},\ and\ \citenamefont {Niu}}]{Gao14}%
  \BibitemOpen
  \bibfield  {author} {\bibinfo {author} {\bibfnamefont {Y.}~\bibnamefont
  {Gao}}, \bibinfo {author} {\bibfnamefont {S.~A.}\ \bibnamefont {Yang}}, \
  and\ \bibinfo {author} {\bibfnamefont {Q.}~\bibnamefont {Niu}},\ }\href
  {\doibase 10.1103/PhysRevLett.112.166601} {\bibfield  {journal} {\bibinfo
  {journal} {Phys. Rev. Lett.}\ }\textbf {\bibinfo {volume} {112}},\ \bibinfo
  {pages} {166601} (\bibinfo {year} {2014})}\BibitemShut {NoStop}%
\bibitem [{\citenamefont {Carbotte}(2016)}]{Carbotte16Tilt}%
  \BibitemOpen
  \bibfield  {author} {\bibinfo {author} {\bibfnamefont {J.~P.}\ \bibnamefont
  {Carbotte}},\ }\href {\doibase 10.1103/PhysRevB.94.165111} {\bibfield
  {journal} {\bibinfo  {journal} {Phys. Rev. B}\ }\textbf {\bibinfo {volume}
  {94}},\ \bibinfo {pages} {165111} (\bibinfo {year} {2016})}\BibitemShut
  {NoStop}%
\bibitem [{\citenamefont {Ivchenko}\ and\ \citenamefont {Pikus}(1975)}]{IP75}%
  \BibitemOpen
  \bibfield  {author} {\bibinfo {author} {\bibfnamefont {E.~L.}\ \bibnamefont
  {Ivchenko}}\ and\ \bibinfo {author} {\bibfnamefont {G.~E.}\ \bibnamefont
  {Pikus}},\ }\href@noop {} {\bibfield  {journal} {\bibinfo  {journal} {Sov.
  Phys. Solid State}\ }\textbf {\bibinfo {volume} {16}},\ \bibinfo {pages}
  {1261} (\bibinfo {year} {1975})}\BibitemShut {NoStop}%
\bibitem [{\citenamefont {Tsirkin}\ \emph {et~al.}(2018)\citenamefont
  {Tsirkin}, \citenamefont {Puente},\ and\ \citenamefont {Souza}}]{SouzaTe}%
  \BibitemOpen
  \bibfield  {author} {\bibinfo {author} {\bibfnamefont {S.~S.}\ \bibnamefont
  {Tsirkin}}, \bibinfo {author} {\bibfnamefont {P.~A.}\ \bibnamefont {Puente}},
  \ and\ \bibinfo {author} {\bibfnamefont {I.}~\bibnamefont {Souza}},\ }\href
  {\doibase 10.1103/PhysRevB.97.035158} {\bibfield  {journal} {\bibinfo
  {journal} {Phys. Rev. B}\ }\textbf {\bibinfo {volume} {97}},\ \bibinfo
  {pages} {035158} (\bibinfo {year} {2018})}\BibitemShut {NoStop}%
\bibitem [{\citenamefont {Lv}\ \emph {et~al.}(2015{\natexlab{b}})\citenamefont
  {Lv}, \citenamefont {Weng}, \citenamefont {Fu}, \citenamefont {Wang},
  \citenamefont {Miao}, \citenamefont {Ma}, \citenamefont {Richard},
  \citenamefont {Huang}, \citenamefont {Zhao}, \citenamefont {Chen},
  \citenamefont {Fang}, \citenamefont {Dai}, \citenamefont {Qian},\ and\
  \citenamefont {Ding}}]{TaAsLv}%
  \BibitemOpen
  \bibfield  {author} {\bibinfo {author} {\bibfnamefont {B.~Q.}\ \bibnamefont
  {Lv}}, \bibinfo {author} {\bibfnamefont {H.~M.}\ \bibnamefont {Weng}},
  \bibinfo {author} {\bibfnamefont {B.~B.}\ \bibnamefont {Fu}}, \bibinfo
  {author} {\bibfnamefont {X.~P.}\ \bibnamefont {Wang}}, \bibinfo {author}
  {\bibfnamefont {H.}~\bibnamefont {Miao}}, \bibinfo {author} {\bibfnamefont
  {J.}~\bibnamefont {Ma}}, \bibinfo {author} {\bibfnamefont {P.}~\bibnamefont
  {Richard}}, \bibinfo {author} {\bibfnamefont {X.~C.}\ \bibnamefont {Huang}},
  \bibinfo {author} {\bibfnamefont {L.~X.}\ \bibnamefont {Zhao}}, \bibinfo
  {author} {\bibfnamefont {G.~F.}\ \bibnamefont {Chen}}, \bibinfo {author}
  {\bibfnamefont {Z.}~\bibnamefont {Fang}}, \bibinfo {author} {\bibfnamefont
  {X.}~\bibnamefont {Dai}}, \bibinfo {author} {\bibfnamefont {T.}~\bibnamefont
  {Qian}}, \ and\ \bibinfo {author} {\bibfnamefont {H.}~\bibnamefont {Ding}},\
  }\href {\doibase 10.1103/PhysRevX.5.031013} {\bibfield  {journal} {\bibinfo
  {journal} {Phys. Rev. X}\ }\textbf {\bibinfo {volume} {5}},\ \bibinfo {pages}
  {031013} (\bibinfo {year} {2015}{\natexlab{b}})}\BibitemShut {NoStop}%
\bibitem [{\citenamefont {Sturman}\ and\ \citenamefont
  {Fridkin}(1992)}]{SturmanBook}%
  \BibitemOpen
  \bibfield  {author} {\bibinfo {author} {\bibfnamefont {B.}~\bibnamefont
  {Sturman}}\ and\ \bibinfo {author} {\bibfnamefont {V.}~\bibnamefont
  {Fridkin}},\ }\href@noop {} {\emph {\bibinfo {title} {Photovoltaic and
  Photo-refractive Effects in Noncentrosymmetric Materials}}}\ (\bibinfo
  {publisher} {Gordon and Breach},\ \bibinfo {year} {1992})\BibitemShut
  {NoStop}%
\bibitem [{\citenamefont {Koepernik}\ and\ \citenamefont
  {Eschrig}(1999)}]{koepernik1999full}%
  \BibitemOpen
  \bibfield  {author} {\bibinfo {author} {\bibfnamefont {K.}~\bibnamefont
  {Koepernik}}\ and\ \bibinfo {author} {\bibfnamefont {H.}~\bibnamefont
  {Eschrig}},\ }\href@noop {} {\bibfield  {journal} {\bibinfo  {journal}
  {Physical Review B}\ }\textbf {\bibinfo {volume} {59}},\ \bibinfo {pages}
  {1743} (\bibinfo {year} {1999})}\BibitemShut {NoStop}%
\bibitem [{\citenamefont {Perdew}\ \emph {et~al.}(1996)\citenamefont {Perdew},
  \citenamefont {Burke},\ and\ \citenamefont {Ernzerhof}}]{perdew1996}%
  \BibitemOpen
  \bibfield  {author} {\bibinfo {author} {\bibfnamefont {J.~P.}\ \bibnamefont
  {Perdew}}, \bibinfo {author} {\bibfnamefont {K.}~\bibnamefont {Burke}}, \
  and\ \bibinfo {author} {\bibfnamefont {M.}~\bibnamefont {Ernzerhof}},\
  }\href@noop {} {\bibfield  {journal} {\bibinfo  {journal} {Phys. Rev. Lett.}\
  }\textbf {\bibinfo {volume} {77}},\ \bibinfo {pages} {3865} (\bibinfo {year}
  {1996})}\BibitemShut {NoStop}%
\end{thebibliography}
\end{document}